\documentclass[aps,reprint, showkeys, nofootinbib,superscriptaddress,pre]{revtex4-2}
\usepackage{CJK}
\usepackage[utf8]{inputenc}
\usepackage{subcaption}
\usepackage{amsmath}
\usepackage{breqn}
\usepackage{colortbl}
\usepackage{nicefrac}
\usepackage{xcolor} 
\usepackage{graphicx}
\usepackage{dcolumn}
\usepackage{bm}
\usepackage[normalem]{ulem} 

\newcommand\fsout{\bgroup\markoverwith{\textcolor{orange}{\rule[0.5ex]{1pt}{1pt}}}\ULon} 

\newcommand\lsout{\bgroup\markoverwith{\textcolor{purple}{\rule[0.5ex]{1pt}{1pt}}}\ULon} 

\newcommand\gsout{\bgroup\markoverwith{\textcolor{teal}{\rule[0.5ex]{1pt}{1pt}}}\ULon} 
\newcommand{\gabi}{\color{teal}} 

\newcommand\msout{\bgroup\markoverwith{\textcolor{red}{\rule[0.5ex]{1pt}{1pt}}}\ULon} 

\makeatletter
\let\cat@comma@active\@empty
\makeatother

\begin{document}

	\title{Shannon information criterion for low-high diversity transition in Moran and Voter models}
	
	\author{Gabriella Dantas Franco}
	\affiliation{
		Universidade Estadual de Campinas, 
		Departamento de F\'isica da Mat\'eria Condensada, 
		Instituto de F\'isica  `Gleb Wataghin', 
		13083-970 -- Campinas - SP  Brazil}
	
	\author{Flavia Maria Darcie Marquitti}
	\affiliation{
		Universidade Estadual de Campinas, 
		Departamento de F\'isica da Mat\'eria Condensada, 
		Instituto de F\'isica  `Gleb Wataghin', 
		13083-970 -- Campinas - SP  Brazil}
	\affiliation{
		Universidade Estadual de Campinas, 
		Instituto de Biologia, 
		13083-862 -- Campinas - SP  Brazil}
	
	\author{Lucas D. Fernandes}
	\affiliation{Department of Life Sciences, 
		Imperial College London - Silwood Park, 
		SL5 7PY - Ascot - Berkshire -- UK}
	
	\author{Dan Braha}
	\affiliation{New England Complex Systems Institute, 
		Cambridge - MA -- USA}
	\affiliation{University of Massachusetts,
		Dartmouth - MA -- USA}

	\author{Marcus Aloizio Martinez de Aguiar}
	\email{aguiar@ifi.unicamp.br}
	\affiliation{
		Universidade Estadual de Campinas, 
		Departamento de F\'isica da Mat\'eria Condensada, 
		Instituto de F\'isica  `Gleb Wataghin', 
		13083-970 -- Campinas - SP  Brazil}
	
	\date{\today}
	
	\begin{abstract}
		
		Mutation and drift play opposite roles in genetics. While mutation creates diversity, drift can cause gene variants to disappear, especially when they are rare. In the absence of natural selection and migration, the balance between the drift and mutation in a well-mixed population defines its diversity. The Moran model captures the effects of these two evolutionary forces and has a counterpart in social dynamics, known as the Voter model with external opinion influencers. Two extreme outcomes of the Voter model dynamics are consensus and coexistence of opinions, which correspond to low and high diversity in the Moran model. Here we use a Shannon's information-theoretic approach to characterize the smooth transition between the states of consensus and coexistence of opinions in the Voter model. Mapping the Moran into the Voter model we extend the results to the mutation-drift balance and characterize the transition between low and high diversity in finite populations. Describing the population as a network of connected individuals we show that the transition between the two regimes depends on the network topology of the population and on the possible asymmetries in the mutation rates.
		
	\end{abstract}
	
	
	
	\maketitle
	
	\section{Introduction}
	\label{sec::intro}

	Consensus dynamics in systems composed by multiple agents is a fundamental aspect in many areas of social and biological sciences  \cite{becchetti2020consensus,villaverde2015consensus}. 
	Understanding the structural and dynamical conditions that allow consensus to be established has both theoretical \cite{ben2003unity,carletti2006make,furioli2017fokker} and applied implications \cite{braha2017}. Interest in the subject 
	covers several important phenomena, such as dynamics of cultural elements \cite{vazquez2010agent} and flow of information in society \cite{wu2004information}, epidemic spreading \cite{moreno2002epidemic,barrat2008dynamical}, and animal collective behavior \cite{sumpter2006principles,ballerini2008interaction}. In particular, the problems of allelic drift in population genetics \cite{gillespie2004population} and opinion dynamics in human societies may be treated under this framework and have classically been described with two theoretical foundations: the Moran model and the Voter model.
	
	The Moran model is a birth-death model describing the evolution of allelic frequencies in a haploid population of fixed size and two alleles \cite{moran1958random}. In its classical formulation, at each time step one individual is chosen to die and another is chosen to reproduce, with the offspring copying the allele of the parent. This dynamics can be directly mapped into the Voter model \cite{liggett2012interacting}, with the two alleles playing the role of two opinions or political affiliations, and each individual adopting the opinion of a randomly selected neighbor at each time step. A natural extension of the Moran model includes the possibility of mutations, allowing the allele of the newborn individual to differ from that of its parent. The role played by mutations in the Moran model corresponds to external influencers, or zealots, in the Voter model \cite{mobilia2005voting,chinellato2015dynamical}, which are a set of frozen-state individuals that can have their opinion adopted, or copied, by other individuals but never change their own state. The system dynamics' can then be described directly in terms of master equations \cite{chinellato2015dynamical} or in terms of the rates of change between states \cite{kirman1993ants,alfarano2005estimation}.
	
	In the absence of mutations, the Moran model converges to a complete homogeneous state just due to fluctuations, also called genetic drift. When mutation is bidirectional  fixation of a single allele is prevented. This case characterizes the so-called mutation-drift balance \cite{kimura1983neutral}, which leads to higher genetic diversity within the population  \cite{gillespie2004population}. Correspondingly, in the Voter model the presence of influencers for both candidates prevents the formation of consensus \cite{mobilia2007role,chinellato2015dynamical}. Complete allele fixation in the Moran model and consensus in the Voter model are called absorbing states, where all the individuals have the same state in equilibrium. 
	
	The total number of individuals having the same opinion (allele) defines a macrostate in the Voter (Moran) model. The presence of external influencers leads to non-trivial steady-state probability distributions of macrostates, since consensus formation would be permanently disturbed. Two phases can then be distinguished in both models, depending on the number of influencers or the values of mutation rates: a phase of low diversity, with a strong prevalence of one opinion (or allele), and a phase of high diversity, where both opinions (or alleles) coexist. In both cases the macrostates settle to a steady-state distribution, but with microstates constantly changing. The steady state distribution depends continuously on the number of influencers (or mutation rate) and there is no sharp transition between the two phases. However, when the population is structured by an underlying network of connections (see Methods), the properties of the network affect the parameter region where the smooth transition between the phases occurs. 
	
	The importance of defining and detecting the transition point in these finite systems is apparent in both biological and social contexts. In biological populations, low diversity is associated with monomorphism (allele fixation), whereas high diversity indicates coexistence of different morphs (different alleles). If the only source of genetic variation is mutation, knowledge of the critical mutation rate is important to understand if the population can evolve to a state of high diversity, where natural selection can act, leading, for example, to speciation. Transition to a state of high diversity can be caused not only by an increase in the mutation rate, but also by a restructure of the spatial distribution of its individuals \cite{schneider2016mutation}. Similarly, in the context of the Voter model, the topology of the network of contacts in a social group can affect the outcome of an election, changing it from consensus to coexistence of opinions under fixed external influences. Thus, in order to understand how network structure alters the  transition between the states it is important to have a precise and sensible definition of the transition point, which can be applied to any network topology and levels of external influence.
	
	For well-mixed populations, corresponding to fully connected networks, analytical solutions for the steady-state probability distribution of the Voter model can be obtained \cite{chinellato2015dynamical} and used to  identify the transition point between the high and low diversity phases. In particular, as we move around in the parameter space of the model, the stationary distributions exhibit strikingly different shapes. The high diversity phase is characterized by unimodal distributions with intermediate peaks, whereas the low diversity phase is characterized by unimodal or bimodal distributions with peaks at extreme values of the node state values. In this case, the uniform probability distribution, where all states are equally likely, marks the transition between the two phases  \cite{chinellato2015dynamical,schneider2016mutation}. This transition is obtained when the number of influencers for each opinion is equal to one, the value of which is defined as the transition point of the model. Although an approximation can be obtained for the steady-state distribution in structured populations with more general network topologies \cite{chinellato2015dynamical,schneider2016mutation}, the transition between the two phases is less well-defined analytically compared with fully connected networks, and cannot be determined by direct analysis of the shape of the distribution. 
	
	In the language of statistical physics, low and high diversity states correspond to ordered and disordered phases, respectively. However, because we are dealing with finite systems and smooth crossover between the phases, there is no evident order parameter to mark the point of transition. Here we propose a practical approach for analyzing the transition between high and low diversity states in the Moran or the Voter models using the Shannon entropy. In the context of Information Theory, the entropy expresses the inherent uncertainty associated with the occurrence of given states in the system. Accordingly, we characterize the transition point as the parameter values that maximize the uncertainty around the macrostates and, therefore, the Shannon entropy. The method is consistent with the fully mixed population case, and is sufficiently flexible to characterize the transition in general structured populations with symmetric or asymmetric mutation rates (analogously, number of influencers in the Voter model). We develop a mean-field approximation for the transition point and compare its accuracy with results -- obtained by the information-theoretic method -- for ring and lattice networks. Finally, we analyze the effect of network degree, asymmetries in mutation rates (or external influencers), and network randomness on the smooth low-high diversity transition.

	\section{Methods}
	\label{sec::Methods}
	
	\subsection{Moran and Voter models}
	\label{subsec::moran_voter}
	
	In order to describe the transition between high and low diversity in the Moran model we explore its analogy with the Voter model with external influencers. The Moran model describes a population where each individual $i$ is defined by a biallelic gene $x_i$, with $x_i \in\{0,1\}$ representing two possible alleles \cite{moran1958random,cannings_1974,ewens2012mathematical,nowak2006evolutionary}. Individuals are haploid and the population has a finite and fixed size $N$. Here, we use a modified version of the Moran model that allows for sexual reproduction and mutation \cite{schneider2016mutation}, with a structured population represented as a network linking potential sexual partners. At each step, a focal parent is selected to mate with a sexual partner. The focal and the partner individuals can mate only if they are linked in a network of interactions, defined by an adjacency matrix ${\bf A}$ whose elements $A_{ij}$ are assigned $1$ if the individuals $i$ and $j$ are connected (reads as ``$i$ can reproduce with $j$'') and $0$ otherwise. The offspring inherits the allele of one of the parents with equal probability. After reproduction there is a chance $\mu_-$ that the offspring gene mutates from allele 0 to 1 and a chance $\mu_+$ that the offspring gene mutates from allele 1 to 0. Finally, the focal parent is replaced by the offspring.
	
	\begin{figure}[!htpb]
		\centering
		\includegraphics[width=1\linewidth]{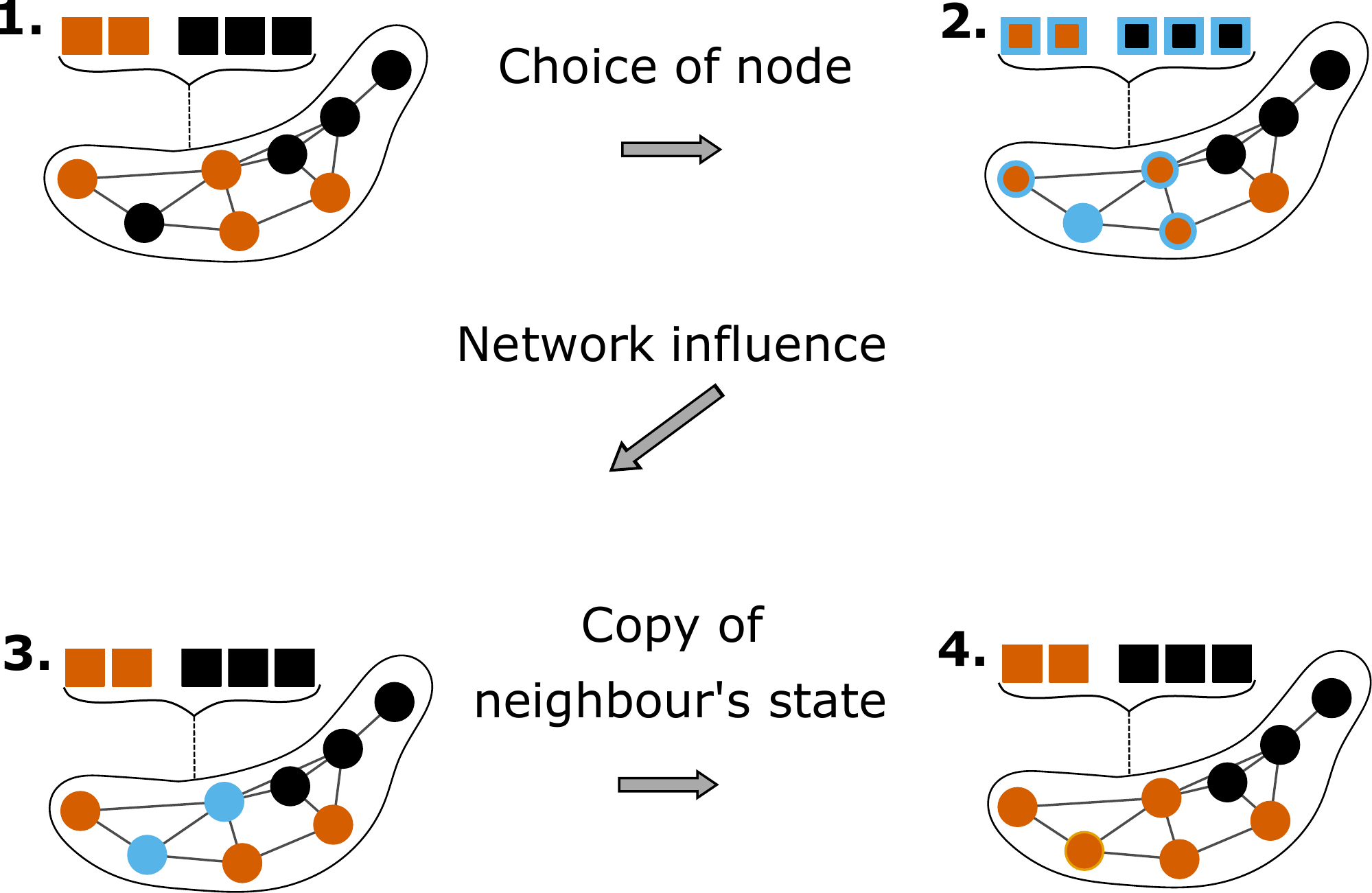}
		\caption{(color online) The Voter model with external influencers. (1) Voters are represented by circles in a network of social contacts, squares are external influencers, and colors (shades of gray) represent opinions. (2) A focal individual is randomly selected to be updated (blue, or light grey, circle). Its neighbors are shown with a thick blue (light grey) border. (3) The focal individual chooses a random neighbor to copy. (4) The state of the focal individual is updated. }
		\label{fig::dynamics}
	\end{figure}
	
	In the Voter model, individuals decide between two candidates in an election (represented by candidates 0 and 1), while being influenced in their vote opinion by their peers \cite{kirman1993ants,alfarano2005estimation,liggett2012interacting,braha2017}. A population of voters is represented by $N$ nodes in a network of social connections, in which first neighbors can communicate and influence each other's votes. In addition, external nodes with fixed opinions may also be present, and they are connected to every node in the network, influencing all the individuals. This simulates the effect of strong political supporters or political propaganda \cite{harmon2015anticipating,khalil2018zealots}. The number of nodes with fixed opinions are defined as $N_0$ (supporting candidate 0) and $N_1$ (supporting candidate 1). Here we will refer to the degree of node $i$ -- $k_i$ -- as the the number of connections with other voters, excluding the external influencers. 
	
	At each time step, one of the voters is randomly chosen (step 2 in Fig. \ref{fig::dynamics}), keeping its opinion with probability $p$ or copying the opinion of one of its neighbors with probability $(1-p)$ (step 3 in Fig. \ref{fig::dynamics}). The opinion is copied  from one of the neighbors in the network or one of the external influencers (fixed nodes). 
	
	Both models can be described by a Markov chain \cite{todorovic2012introduction} with analogous master equations describing the probability $P_{t+1}({\bf x})$ of having a microstate ${\bf x} = (x_1, x_2, \dots, x_N)$ at time  $t+1$ given the probabilities at time $t$ (see Appendix \ref{app:maseqn}). For regular networks, where the degree of a node $i$ is $k_i=k$ for all nodes, there is an exact correspondence between the master equations, allowing for a map between the parameters of the two models. In such case, the mapping is given by:
	
	\begin{equation}
		\left\{
		\begin{array}{ll}
			N_1 &= {\displaystyle\frac{2\mu_- k}{1-2\bar{\mu}}} \\ \\
			N_0 &= {\displaystyle\frac{2\mu_+ k}{1-2\bar{\mu}}}
		\end{array} \right. \qquad \mbox{or} \qquad
		\left\{
		\begin{array}{ll}
			\mu_+ &= {\displaystyle\frac{N_0}{N_0+N_1+2k}} \\ \\
			\mu_- &= {\displaystyle\frac{N_1}{N_0+N_1+2k}}
		\end{array} \right. 
		\label{mapping_voter_moran}
	\end{equation}
	and
	\begin{equation}
		p=\frac{1}{2}-\bar{\mu},
		\label{prob_mu_relation}
	\end{equation}

	\noindent where $\bar{\mu} = (\mu_{+}+\mu_{-})/2$ (see Appendix \ref{app:maseqn} for complete derivation). The parameter $p$ is only relevant for the equilibration time, and has no effect on the stationary probability distribution.  When $\mu_{+}\neq\mu_{-}$ (or $N_0\neq N_1$), the steady-state distribution is asymmetric {\gabi \cite{alfarano2005estimation}} and it is convenient to define the variable $\Delta=\mu_{+}-\mu_{-}$ as a measure of the asymmetry (see subsection \ref{subsec::asym}).
	
	The dynamics for both Voter and Moran models can be shown to reach a stationary probability distribution which is analytically solvable for the case of a fully connected network \cite{chinellato2015dynamical}, when all dynamical nodes are first neighbors ($k_i = N-1$). The symmetry of such topology implies that every microstate ${\bf x}$ corresponding to the macrostate with $m$ nodes at state $1$ has the same probability $\binom{N}{m}$. The corresponding stationary probability distribution for the macrostate $m$ is given by $\rho(m,N_0,N_1)$ (see Eq. \ref{probn} in Appendix \ref{app:maseqn}). 
	
	The analytical solution for $\rho(m,N_0,N_1)$ allows the extension of $N_0$ and $N_1$ to real numbers. Values $0<N_0,N_1<1$ can be interpreted as weak perturbations. Figure  \ref{fig::FC_plots}(a) shows examples of stationary distributions for symmetric numbers of external influencers -- $N_0=N_1$. There are three distinct scenarios:
	
	\begin{itemize}
		\item For sufficiently small $N_0$ and $N_1$ ($N_0, N_1 \ll 1$ -- black curve in Fig. \ref{fig::FC_plots}(a), the population tends to a consensus, where the stationary distribution has high probability of having all the voters for candidate $0$ ($m=0$) or for candidate $1$ ($m= N$), and small chances of mixed opinions in the population ($0< m < N$).
		
		\item For sufficiently large $N_0$ and $N_1$ ($N_0, N_1 \gg 1$ --  blue curve in Fig. \ref{fig::FC_plots}(a), the population is highly affected by external influencers resulting in frequent opinion shifts by voters. In this scenario, the highest probability is for a population with coexisting opinions.
		
		\item For $N_0=N_1=1$ (orange curve in Fig. \ref{fig::FC_plots}(a)), we obtain the uniform distribution $\rho(m, N_0, N_1)=\frac{1}{N+1}$ for all values of $N$, i.e. $N_0=N_1=1$ is the transition value of this model. In this case, all macrostates are equally likely and the system executes a random walk through the state space. The value $T_c\equiv N_0=N_1=1$ marks the transition between low and high opinion diversity states independently of network size. 
	\end{itemize}
	
	In the language of the Moran model the first two regimes correspond, respectively, to strong drift or balance between drift and mutations \cite{gillespie2004population}. The transition between the low diversity (consensus) and high diversity (coexistence of opinions) phases is well defined and obtained for $N_0=N_1=1$, which corresponds to a uniform distribution (orange curve in Fig. \ref{fig::FC_plots}(a)). However, for the asymmetric scenario -- $N_0 \neq N_1$, Fig. \ref{fig::rho_shannon_asym}(a) -- it is not clear how to define the transition point even though the analytical solution for the stationary probability distribution is exactly obtained (see Eq. \eqref{probn}). Indeed, unlike the symmetric case where the uniform distribution provides a clear separation of the low and high diversity phases, the stationary distribution for the asymmetric case is never flat in the transition region. To this end, we propose in the next section an information-theoretic definition of the transition point, which can be applied to any network structure with arbitrary external influence parameters.
	
	\subsection{Characterizing the low-high diversity transition via Shannon Entropy}\label{subsec::transition}
	
	Diversity is a central concept in ecology and the social sciences. Several indices have been proposed to measure such complex notion with a single number that could be compared across different communities. Among these indices, the Simpson, the Shannon and Renyi entropies and Hill numbers are the most commonly employed \cite{daly2018ecological}. Here we focus on the Shannon entropy for its connection with physical systems and for being independent of tuning parameters. 
	
	The Shannon entropy \cite{mackay2003information} corresponding to the stationary probability distribution of macrostates is given by 
	\begin{equation}
		S(N_0,N_1)=-\sum_{m=0}^N \rho(m,N_0,N_1)\log_2[\rho(m,N_0,N_1)]
		\label{shannon}
	\end{equation}
	where we omit the dependence of $S$ and $\rho$ on $N$. 
	
\begin{figure}[!htpb]
		\centering
		\includegraphics[width=0.45\textwidth]{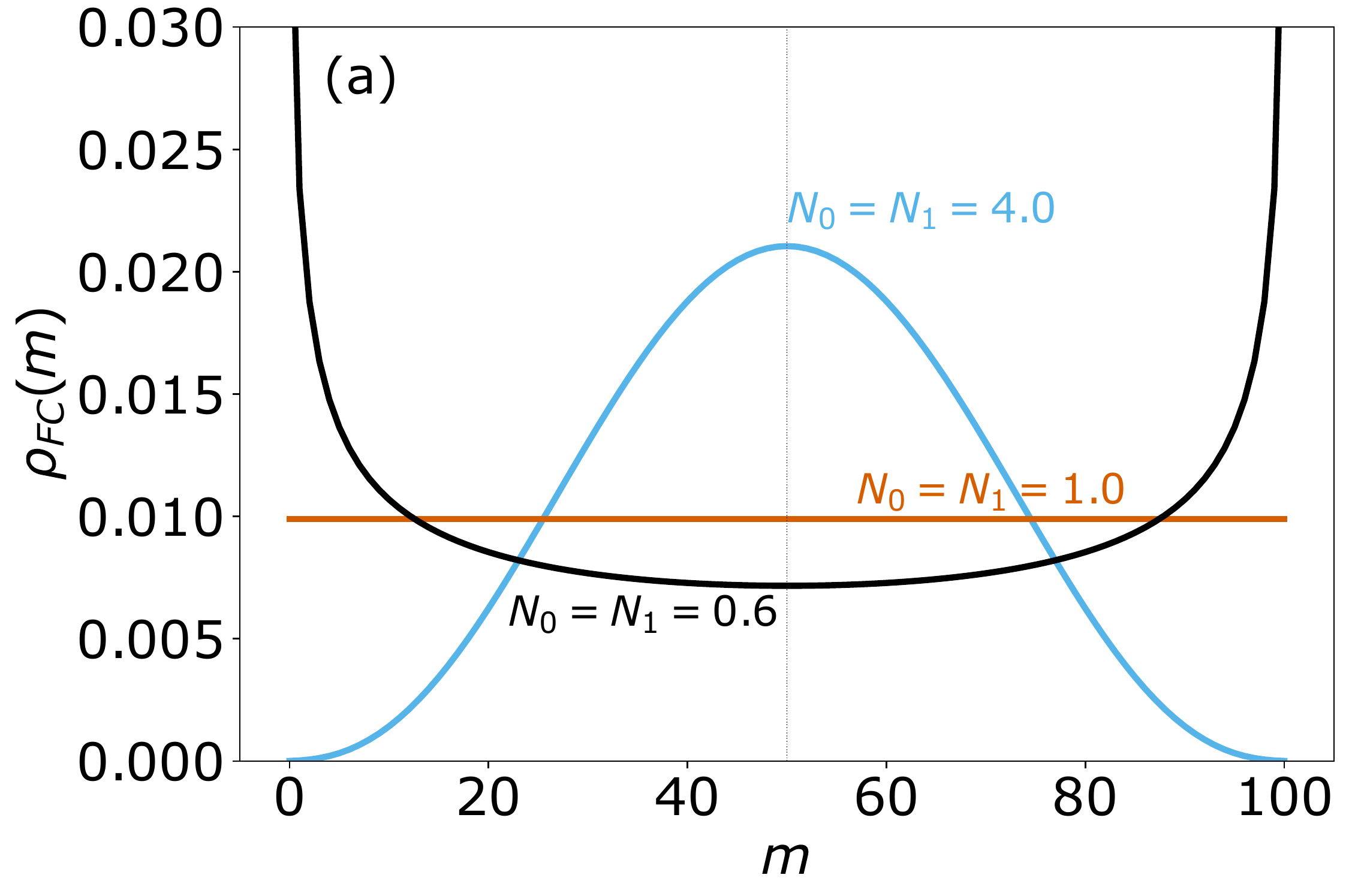}
		\includegraphics[width=0.45\textwidth]{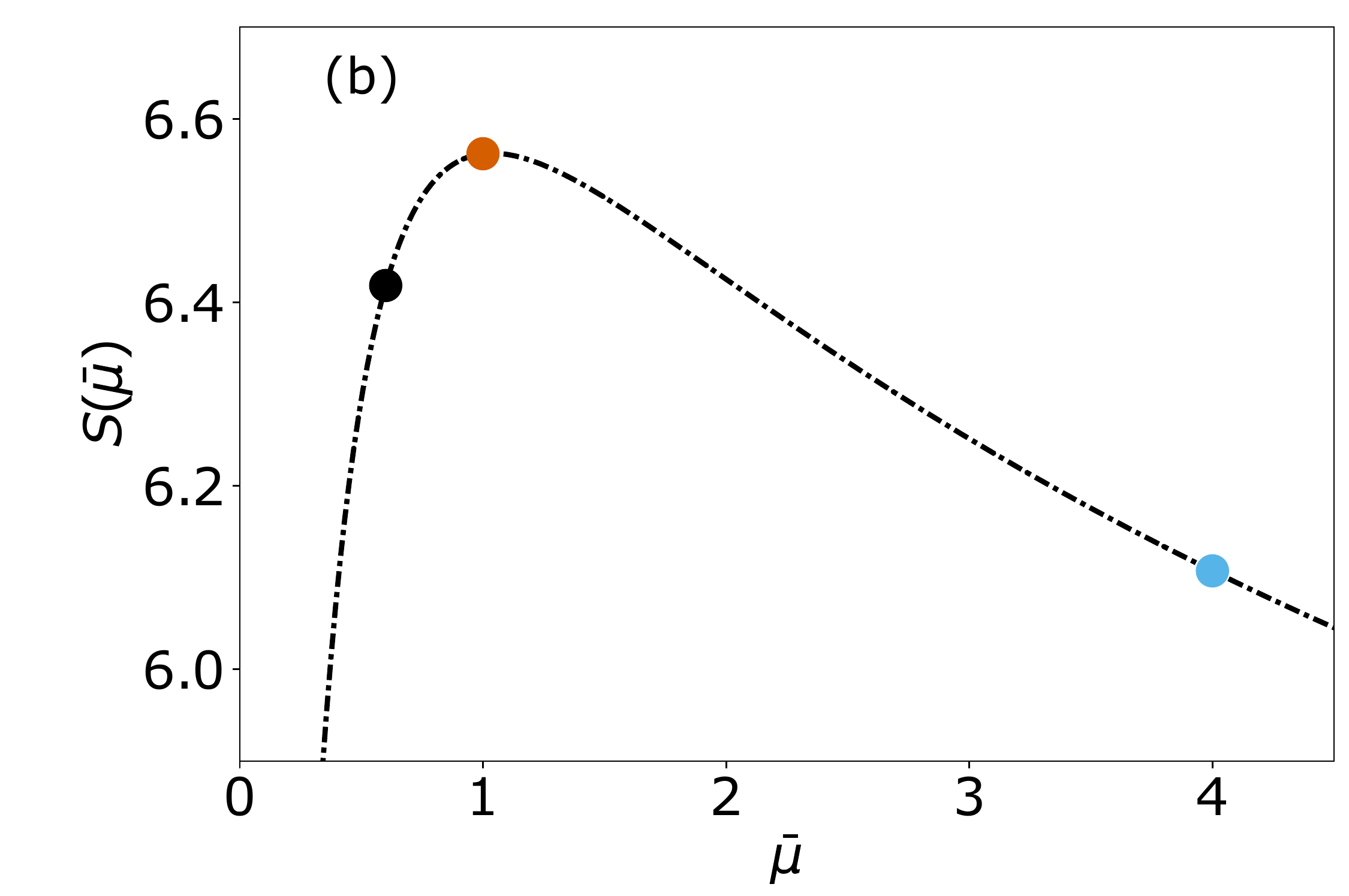}
		\caption{(color online) (a) Stationary probability distributions of macrostates for $T=4.0$ (blue/ light gray), $T=1.0$ (orange/dark gray) and $T=0.6$ (black) and; (b) Shannon entropy as a function of $T$ for a fully connected network with $N=100$ nodes and symmetric external influences ($T = N_0 = N_1$). Colored points in the Shannon entropy function (b) correspond to the curves on panel (a).}
		\label{fig::FC_plots}
	\end{figure}

	To motivate our definition of the high-low diversity transition point, we first consider the special symmetric case ($N_0=N_1$). In this case, the entropy $S(N_0, N_1)$ decreases for larger values of the external influences $N_0$ and $N_1$. Specifically, we obtain that $\rho(m, N_0, N_1) \rightarrow \delta_{m,N/2}$ and $S(N_0,N_0) \rightarrow 0$ in the limit $N_0, N_1\rightarrow\infty$ (see Fig. \ref{fig::FC_plots}(b) and Eq. \ref{probn}). In this limit case, the individuals are distributed evenly between the two alleles (opinions) and the highest genetic diversity  (opinion plurality) within the population is attained. 
	
	The entropy $S(N_0, N_1)$ also decreases as the values of the external influences $N_0$ and $N_1$ decrease. In particular, the distribution peaks at the extremes in the limit $N_0, N_1\rightarrow 0$, tending to $(\delta_{m,0} + \delta_{m,N})/2$ with entropy $S(N_0,N_0) \rightarrow 1$. In this limit case, the individuals are characterized by a single allele (opinion), and the population reaches its lowest diversity (consensus).
	
	At the transition point $N_0 = N_1 = 1$ (see \ref{subsec::moran_voter}), the distribution becomes flat and the entropy attains its maximum value of $\log_2(N + 1)$. The maximum entropy thus interpolates between the high and low diversity phases.
	
	The above considerations lead us to define the transition point for general networks with arbitrary number of external influencers as {\it the set of parameters whose corresponding stationary distribution maximizes the Shannon entropy}. For networks with asymmetric external influences or mutation rates, it is useful to define the total influence $T \equiv (N_0 + N_1)/2$ and the asymmetry parameter $\Delta \equiv N_1-N_0$, analogous to temperature and magnetic field in the Ising model \cite{yang1952spontaneous}. For fixed $\Delta$, the entropy can be computed as a function of $T$, and the transition point $T = T_c(\Delta)$ is determined by maximizing the Shannon entropy. In Section \ref{sec::results}, we examine the utility of this definition of the transition point by extensive numerical simulations. We emphasize that the Shannon entropy based on the distribution of macrostates is very different from the usual Statistical Physics entropy based on microstates. The difference is apparent for high diversity states, which has low entropy in our definition but high entropy if computed with microstates.
	
	\subsection{Mean-field approximation for non-fully connected networks}\label{subsec::mean-field}

	While an analytical solution for the stationary probability distribution can be derived for fully connected (FC) networks (\cite{chinellato2015dynamical}, see Eq. \eqref{probn}), analytical solutions for arbitrary networks are more difficult to obtain. To this end, a mean-field approximation for $\rho(m,N_0,N_1)$ has been proposed \cite{chinellato2015dynamical,de2011moran}. The approximation assumes that macrostate distributions for a given network can be obtained from the analytical result of the fully connected network (Eq. \ref{probn}) by replacing $N_0$ and $N_1$ in the non-fully connected network with effective numbers of external influencers corresponding to a fully connected network, $N_0^{ef}$ and $N_1^{ef}$.
	
	For regular networks, where all nodes have the same degree $k$, the probability $P_i$ that node $i$ copies one of the external fixed nodes (influencers) is given by $P_i(N_0,N_1)=(N_0+N_1)/(k+N_0+N_1)$. We seek a correction to the numbers of external nodes that would match this probability to the same probability corresponding to a fully connected network of the same size. Since for the latter we have $k=N-1$, then:
	
	\begin{align}
		\nonumber \frac{N_0+N_1}{N_0+N_1+k}=\frac{N_0^{ef}+N_1^{ef}}{N_0^{ef}+N_1^{ef}+N-1}
	\end{align}
	
	Assuming that the effective external nodes corresponding to the fully connected network are given by a linear scaling correction, $N_0^{ef}=fN_0$ and $N_1^{ef}=fN_1$, to the external nodes $N_0$ and $N_1$ associated with the regular network we obtain:
	\begin{equation*}
		f=\frac{N-1}{k}.
	\end{equation*}
	Using the above scaling correction, the approximate stationary probability distribution for the k-regular network is obtained by setting $\rho_{k}(m,N,N_0,N_1)=\rho_{FC}(m,N,N_0^{ef},N_1^{ef})=\rho_{FC}(m,N,f N_0,f N_1)$. We note that the same result holds for heterogeneous networks by replacing $k$ in the above equations with the average degree of nodes in the network.
	
	For regular networks with symmetric external influences (i.e., $N_0=N_1$) we have $\rho_{k}(m,N,N_0,N_0)=\rho_{FC}(m,N,N_0^{ef},N_0^{ef})=\rho_{FC}(m,N,f N_0 ,f N_0)$. Since the low-high diversity transition point for fully connected networks with symmetric external influences is $T_c=N_0^{ef}= N_1^{ef}=1$, we obtain the approximate low-high diversity transition point for the regular network case as follows:
	\begin{equation}
		T_c(k)= N_0 = N_1= \frac{1}{f} = \frac{k}{N-1}.    
		\label{eq:mean}
	\end{equation}
	
	Using the correspondence between the Moran and Voter models in Eq.(\ref{mapping_voter_moran}), we find that the critical mutation rate for regular population structures with symmetric mutation rates (i.e., $\mu_+=\mu_-$) is given as follows:

	\begin{equation}
		\mu_c(k) = \mu_+ = \mu_- = \frac{T_c(k)}{2\left[T_c(k) + k\right]}
		\label{eq:mu_trans}
	\end{equation}
	By using the result for $T_c(k)$  given in Eq.(\ref{eq:mean}), we find that the mean-field approximation for the transitional mutation rate is independent of $k$:
	\begin{equation}
		\mu_c(k) = \frac{1}{2N}
		\label{eq:mu_mf}
	\end{equation}
	In the next section, we test the accuracy of these approximations by extensive numerical simulations on both homogeneous and heterogeneous networks. 
	
	\section{Results}\label{sec::results}
	
	In this section, we study the low-high diversity transition by numerical simulations in both regular and random networks. We show that: (1) the transition point depends non-linearly on the degree $k$ of the network; that is, there are important corrections to the mean-field approximation presented in Section \ref{subsec::mean-field}, especially for small $k$;  (2) the specific dependence of the transition point on $k$ also depends on the underlying network topology; (3) for networks with arbitrary number of external influencers or mutation rates, the transition point increases with the extent of asymmetry between the model parameters; and (4) the transition point increases with the level of randomness in the network topology.
	
	\subsection{Numerical Setup}
	\label{subsec::exp-setup}
	
	We first present results for the symmetric case $N_0 = N_1 = T$. We ran simulations of the Voter model with varying $T$ and characterized the transition point $T_c$ as the value that  maximizes the Shannon entropy $S(T)$ for each network type and average degree $k$. The probability $p$ of not changing an opinion was set to $p=0$, since it only affects the equilibration time, and has no effect on the stationary distributions \cite{schneider2016mutation}. The dynamics was initially iterated for an equilibration time of 20,000 steps, after which a sample of macrostates (number of network nodes in state 1) was collected at intervals of 10,000 steps to reduce potential correlation between measurements. A collection of 10,000 sampled macrostates was used to estimate the final stationary distribution $\rho(m,T)$, with which we were able to calculate $S(T)$ (Eq. \ref{shannon}). The overall simulation run time after equilibration was $10^8$ time steps. Repeating the procedure for several values of $T$, we obtained the curve $S(T)$ whose maximizer value provided an estimate for the transition point $T_c$. Smooth curves for $S(T)$ are obtained by averaging over 50 replicas for each value of $T$.
	
	\begin{figure*}[t]
		\centering
		\includegraphics[width=0.45\textwidth]{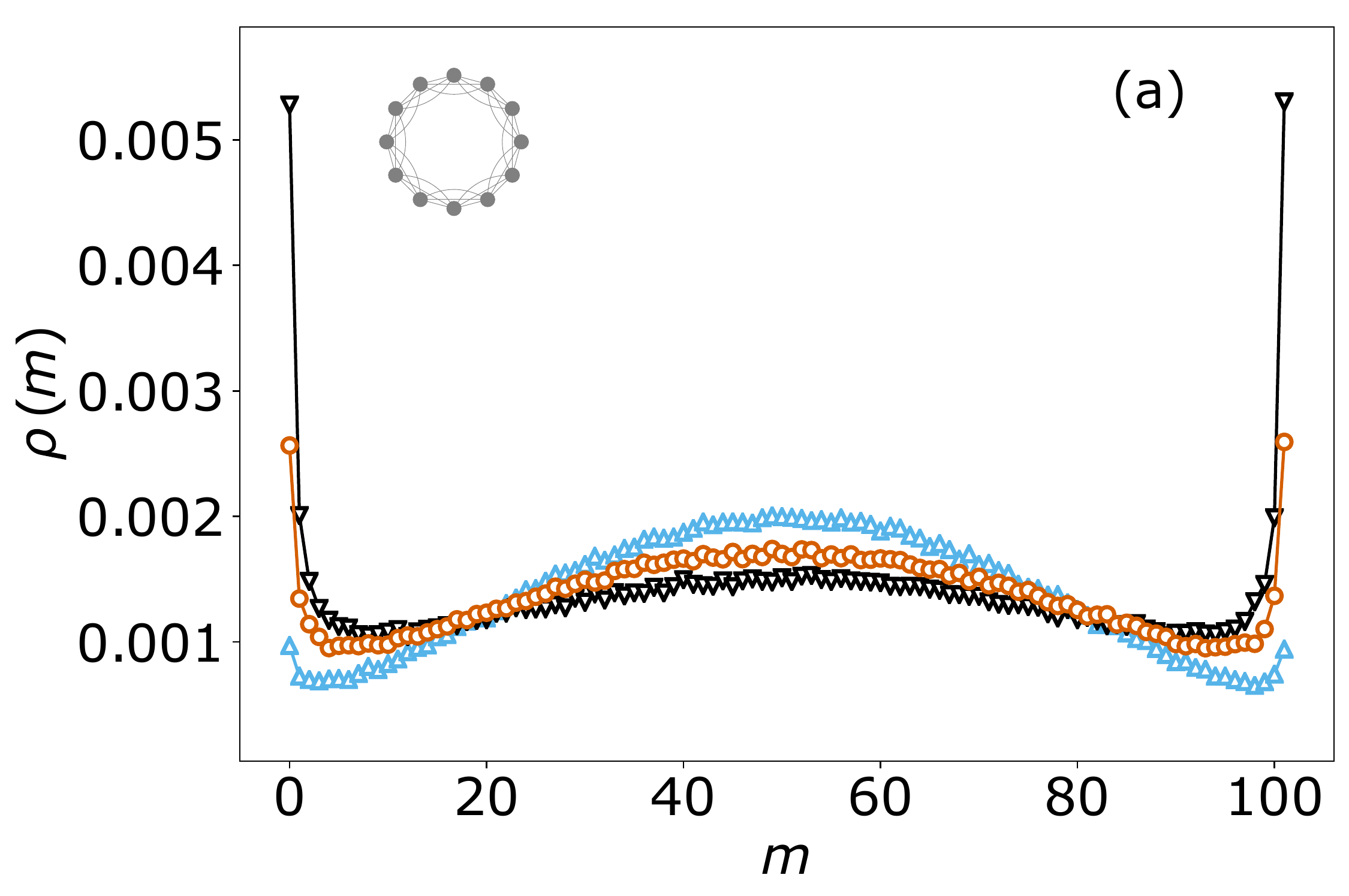}
		\includegraphics[width=0.45\textwidth]{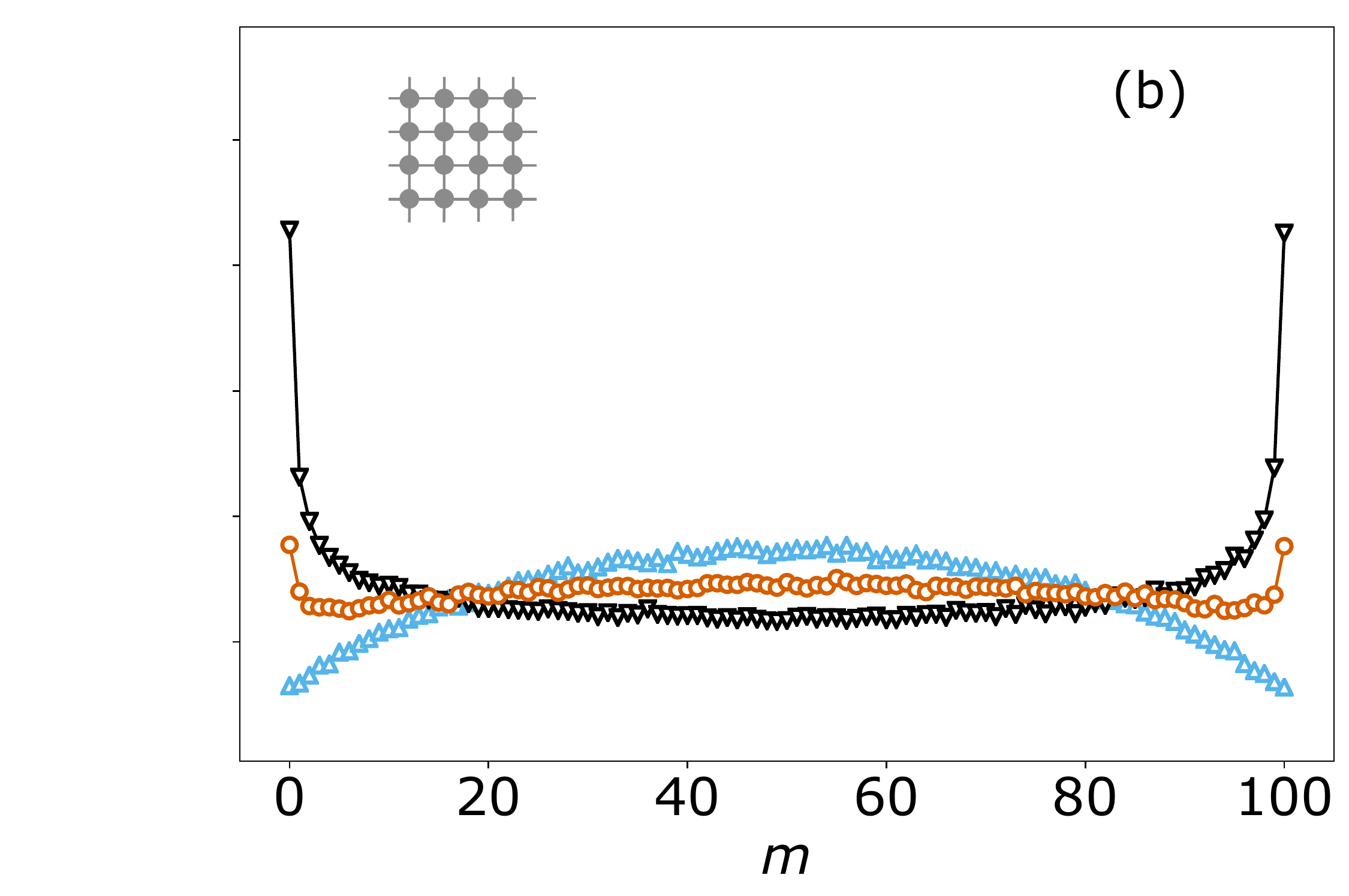}
		\includegraphics[width=0.45\textwidth]{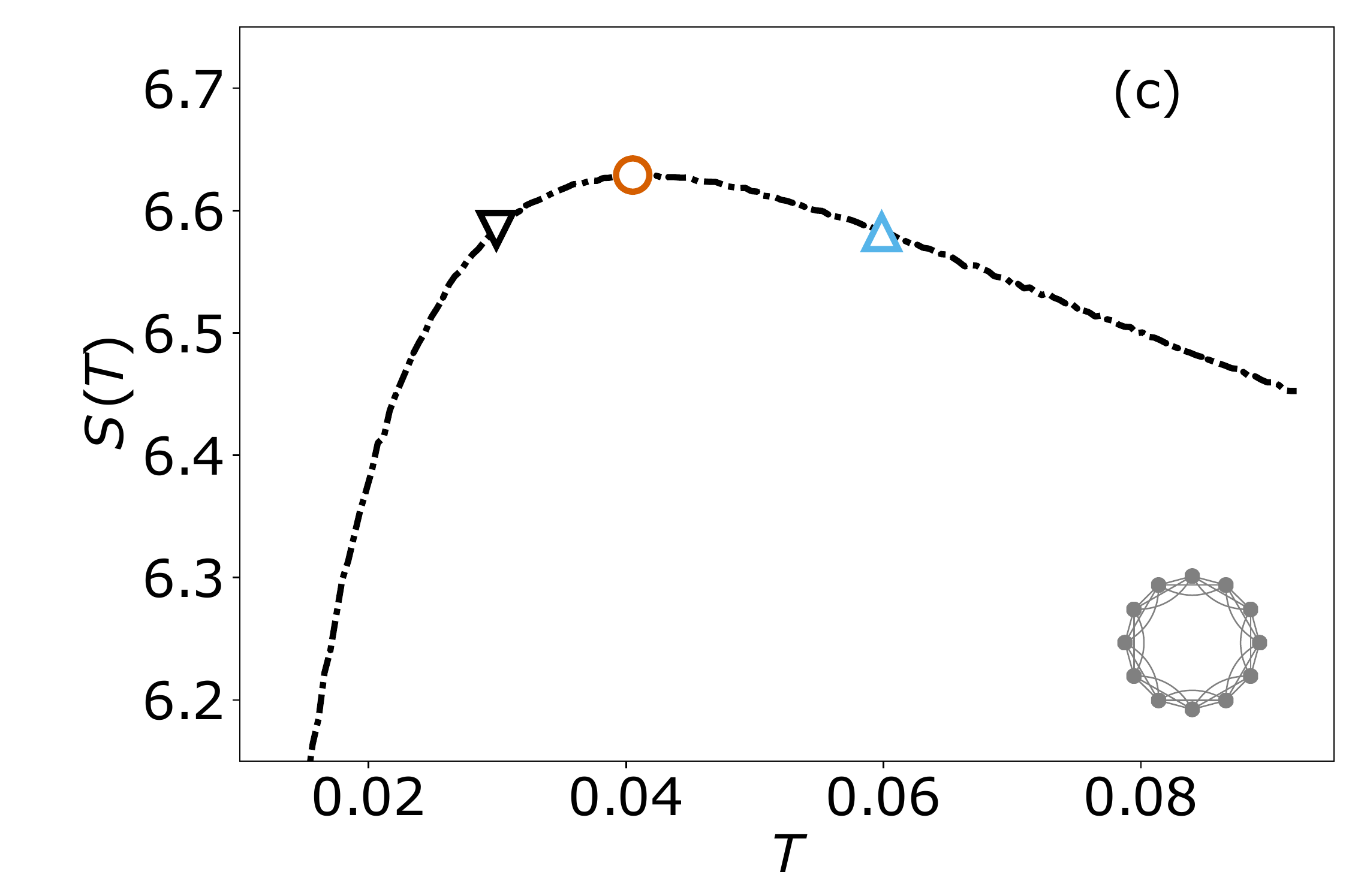}
		\includegraphics[width=0.45\textwidth]{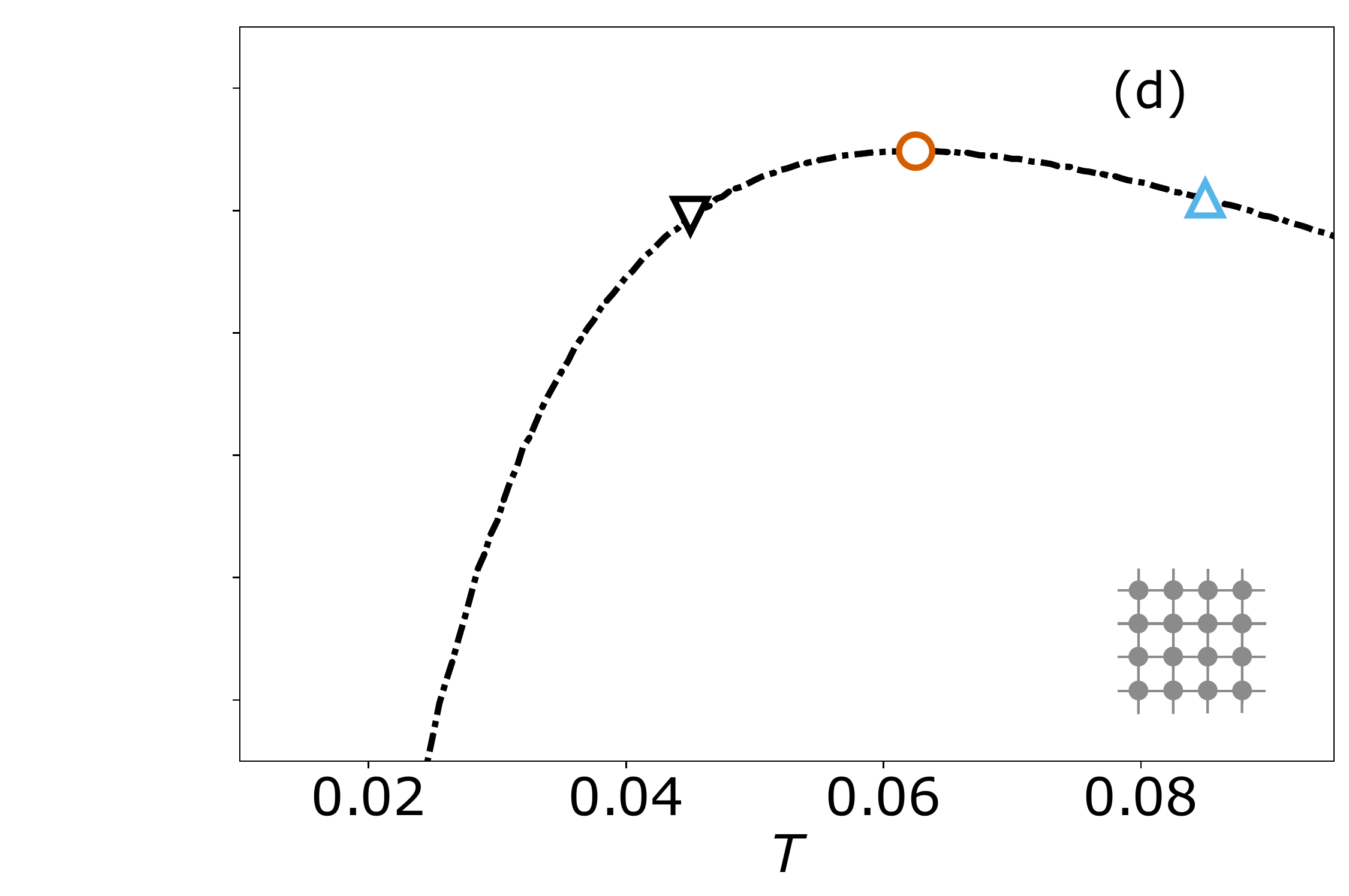}
		\caption{(color online) Stationary probability distributions of macrostates for three selected values of $T$, (a) and (b), and Shannon entropy functions, (c) and (d), for the ring network with $N=101$ nodes, panels (a) and (c), and lattice network with $N=100$ nodes, panels (b) and (d), both with node degree $k=8$. External influences are symmetric ($T = N_0 = N_1$). Colored symbols in the Shannon entropy functions correspond to the curves on the top panels. }
		\label{fig::rho_plts-ring_lattice}
	\end{figure*}

	\subsection{The interplay between network structure and stationary distribution}\label{subsec::strctr-distr}
	
	We investigated how the topology of the network alters the shape of the stationary probability distribution near the transition. Two network topologies with the same node degree ($k = 8$) are explored: a ring network (Figs. \ref{fig::rho_plts-ring_lattice}(a) and \ref{fig::rho_plts-ring_lattice}(c)) and a square lattice network with periodic boundary conditions (Figs. \ref{fig::rho_plts-ring_lattice}(b) and \ref{fig::rho_plts-ring_lattice}(d)). Results for the probability distribution $\rho(m,T)$ and Shannon entropy $S(T)$ are shown on panels (a)-(b) and (c)-(d) of Fig. \ref{fig::rho_plts-ring_lattice}, respectively.
	
	As shown in Fig. \ref{fig::rho_plts-ring_lattice}, different values of $T$ correspond to qualitatively different behaviors of the stationary distributions, varying from low diversity (black) to high diversity (blue triangles). However, in contrast to what is seen for the fully connected case (see Fig. \ref{fig::FC_plots}(a)), the shape of the stationary distribution corresponding to the low-high diversity transition is less well-defined than for fully connected networks (see also the discussion in Section \ref{subsec::nw-strctr-dgr}). The figure suggests, however, that the stationary distribution corresponding to the value of $T$ that maximizes the Shannon entropy interpolates nicely between the low and high diversity phases, indicating the usefulness of the information-theoretic method.

	\subsection{The effect of network structure and degree on the low-high diversity transition}\label{subsec::nw-strctr-dgr}
	
\begin{figure}[!htpb]
		\centering
		\includegraphics[width=0.45\textwidth]{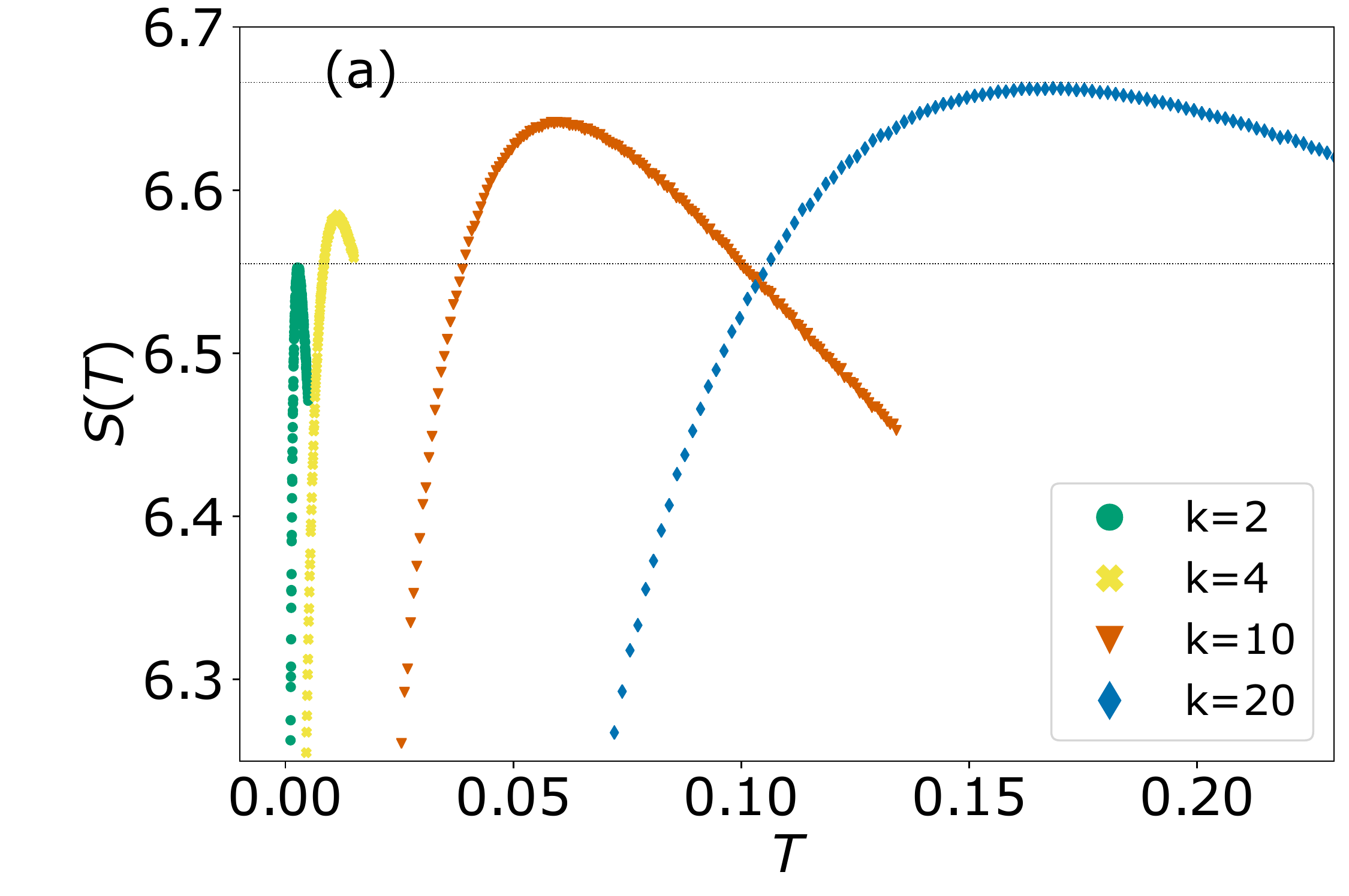}
		\includegraphics[width=0.45\textwidth]{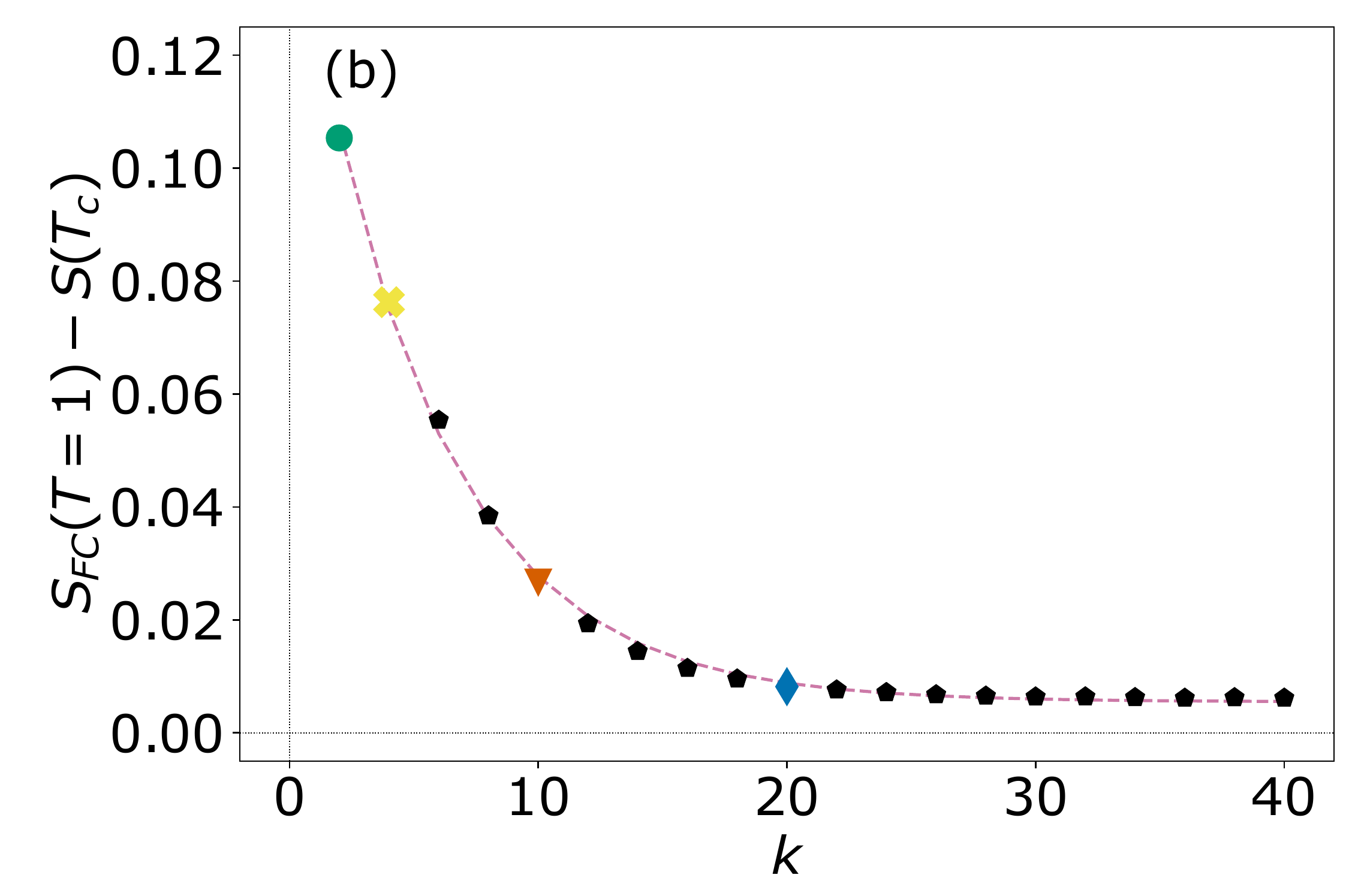}
		\caption{(color online) (a) Comparison between the Shannon entropy functions corresponding to four ring networks with $N=101$ and different node degrees $k$, as a function of the total external influence $T=N_0=N_1$. (b) Difference between the Shannon entropy at the transition point for a fully connected network, $S_{FC}(T=1)$, and the estimated maximum Shannon entropy, $S(T_c)$, for the ring network with varying node degrees $k$. Colored points correspond to the Shannon entropy curves shown on panel (a).}
		\label{fig::results_shannon-comparison}
	\end{figure}

	Here we explore the effect of network degree on the Shannon entropy, and the transition point thereof. Fig. \ref{fig::results_shannon-comparison}(a) shows results for the ring network (with $N = 101$ nodes) with four different network degrees. We see that both transition point $T_c$ and associated maximum entropy value $S(T_c)$ vary with $k$. For the fully connected network ($k=N-1$), the transition occurs when $T_c=1$ and the probability distribution is uniform, corresponding to $S(T_c)=\log_2(N+1)=6.6724 \equiv S_m$, which is the largest possible value for a network with size $N=101$. This implies that any critical value $T_c$ with $S(T_c) < S_m$ corresponds necessarily to a non-uniform equilibrium distribution $\rho(m)$. This departure from uniformity is particularly evident for lower node degrees, where the probability distributions at the transition point ($\rho(m)$ at $T = T_c$) show distinctively non-uniform shapes (see Fig. \ref{fig::rho_plts-ring_lattice}).
	
	To further explore the departure from uniformity at the transition point, we show in Fig. \ref{fig::results_shannon-comparison}(b) the difference between the Shannon entropy at the transition point for a fully connected network, $S_m$, and the estimated maximum Shannon entropy, $S(T_c)$, for a ring network with varying degrees. We show that the difference decreases exponentially with the node degree, reaching the asymptotic value of $0$ as $k$ approaches the value of $N - 1$. This result suggests that distinctively different stationary distributions that substantially deviate from uniformity are detected for lower node degrees ($k < 20$ for the ring network), while rapid convergence towards the uniform distribution are obtained for higher node degrees ($k \approx 30$ for the ring network).
	
\begin{figure*}[ht]
		\centering
		\includegraphics[width=0.39\textwidth]{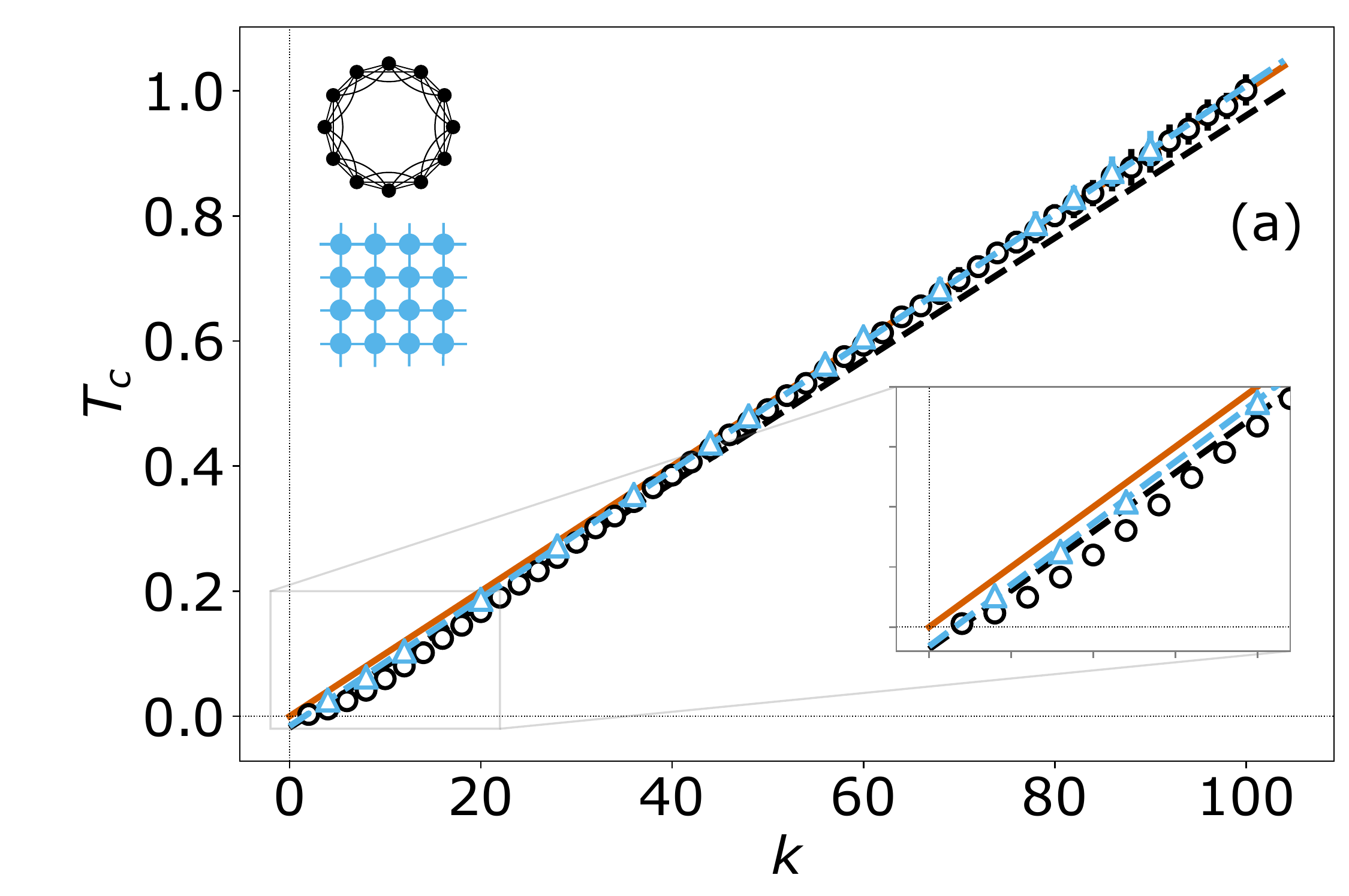}
		\qquad
		\includegraphics[width=0.39\textwidth]{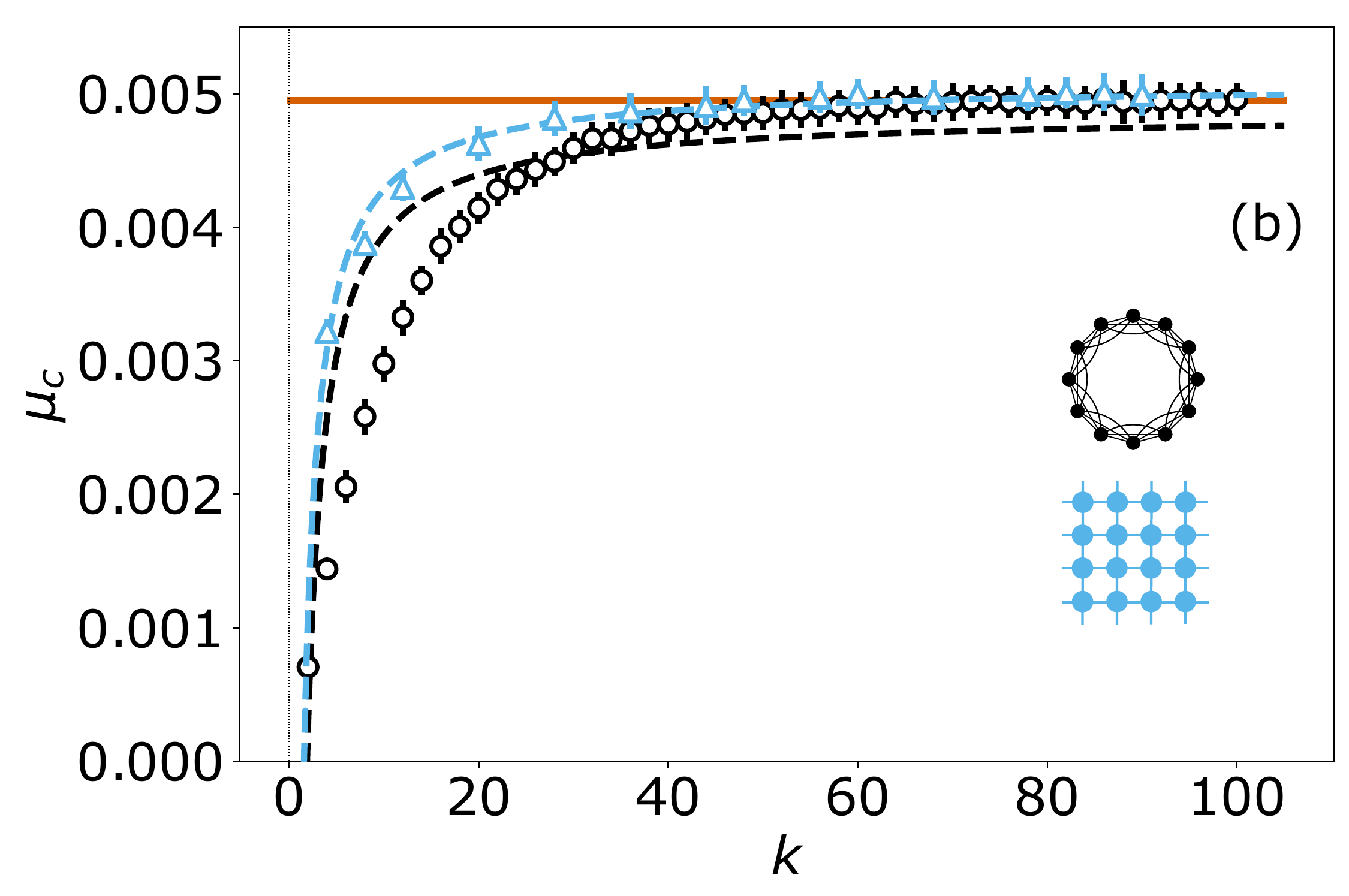}
		\includegraphics[width=0.17\textwidth, trim=0 -3cm 0 0]{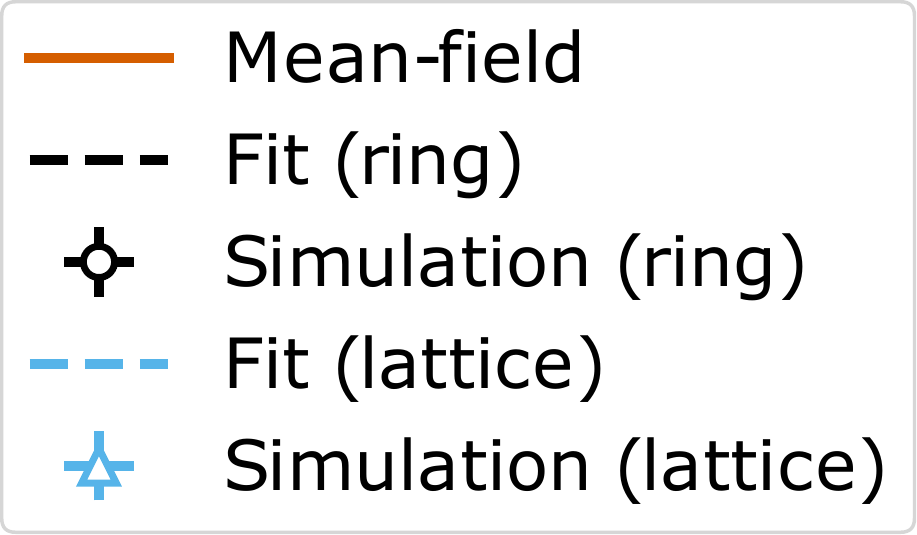}
		\caption{(color online) Transition point, detected via Shannon entropy, as a function of node degree $k$, for the ring ($N=101$) and lattice ($N=100$) networks. Results for the Voter model are shown in panel (a). The corresponding mapped critical mutation rates of the Moran model are shown in panel (b). Error bars on the left correspond to one standard deviation over 50 realizations of the Voter model for each value of $k$ and are plotted for all the points. Error bars for the Moran model are obtained via error propagation using Eq. \ref{eq:mu_trans}. Inset on the left panel shows that the two fitted linear curves have non-zero intercept coefficients.}
		\label{fig::results_degree}
	\end{figure*}

	In order to better understand the coupled effect of network topology and node degree on the transition point, we examine how the node degree affects the transition point in both the ring and lattice networks. Plots of $T_c$ as a function of $k$ for both network topologies are shown in Fig. \ref{fig::results_degree}(a). The corresponding values of the transitional mutation rates for the Moran model ${\mu}_c$ (see Eq. \ref{eq:mu_trans}) as a function of $k$ are shown in Fig. \ref{fig::results_degree}(b).

	We see that for higher node degrees, the numerical values of $T_c$ obtained via the Shannon entropy method show good agreement with the linear mean-field approximation (Eq. \ref{eq:mean}), for both the ring and lattice networks (Fig. \ref{fig::results_degree}(a)). While the deviations between the numerical $T_c$ and the mean-field results for lower values of $k$ seem small (Fig. \ref{fig::results_degree}(a) inset), these deviations are amplified when the Voter model is translated into the corresponding Moran model (see Eq. \ref{eq:mu_trans}). Indeed, the results in Fig. \ref{fig::results_degree}(b) indicate that there is a clear discrepancy between the numerical mutation rates $\mu_c$ for lower values of $k$ and the mean-field approximation $\mu_c=1/2N$ derived in Eq. \ref{eq:mu_mf}.
	
	We obtain approximate curves relating the transitional mutation rate $\mu_c$ to the node degree $k$ by using the linear fit between $T_c$ and $k$  (dashed lines in Fig. \ref{fig::results_degree}(a)) in Eq. \ref{eq:mu_trans} that maps $T_c$ to $\mu_c$. The resulting fits between $\mu_c$ and $k$ are shown in Fig. \ref{fig::results_degree}(b) as the black (ring network) and blue (lattice network) dashed curves. We see that, for lower values of $k$, the fitted curves show a better agreement with the numerical measurements compared with the mean-field prediction. 
	
	\subsection{The effect of asymmetry}
	\label{subsec::asym}

	In addition to the symmetric case analyzed above, we also analyzed the transition point for networks with asymmetric external influences (i.e., $N_0\neq N_1$ or $\mu_+ \neq \mu_-$). In this case, a full characterization of the transition will depend not only on the average value of the external influence, but also on the magnitude of the asymmetry. Without loss of generality, we perform the analysis only for the Moran model with asymmetric mutation rates and fully connected network structure. We define the asymmetry parameter $\Delta = \mu_+-\mu_-$ and study the transition point as we vary $\bar{\mu}=\frac{(\mu_+ + \mu_-)}{2}$ for fixed values of $\Delta$. As above, we apply the Shannon entropy to detect the transition points $\bar{\mu}_c$, which mark the transition between the low and high diversity phases in the Moran model.
	
\begin{figure}[!htpb]
		\centering
		\includegraphics[width=0.45\textwidth]{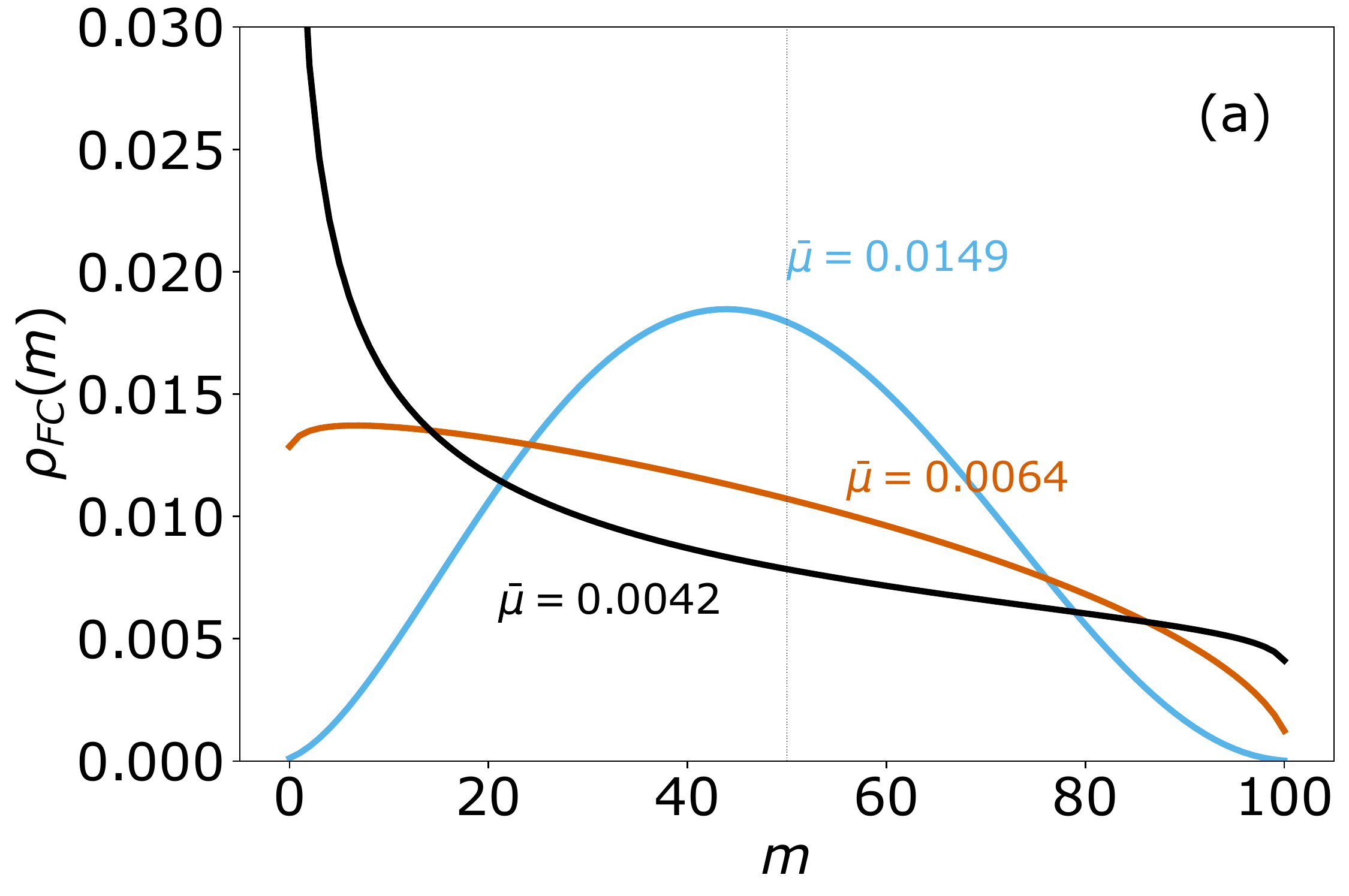}
		
		\includegraphics[width=0.45\textwidth]{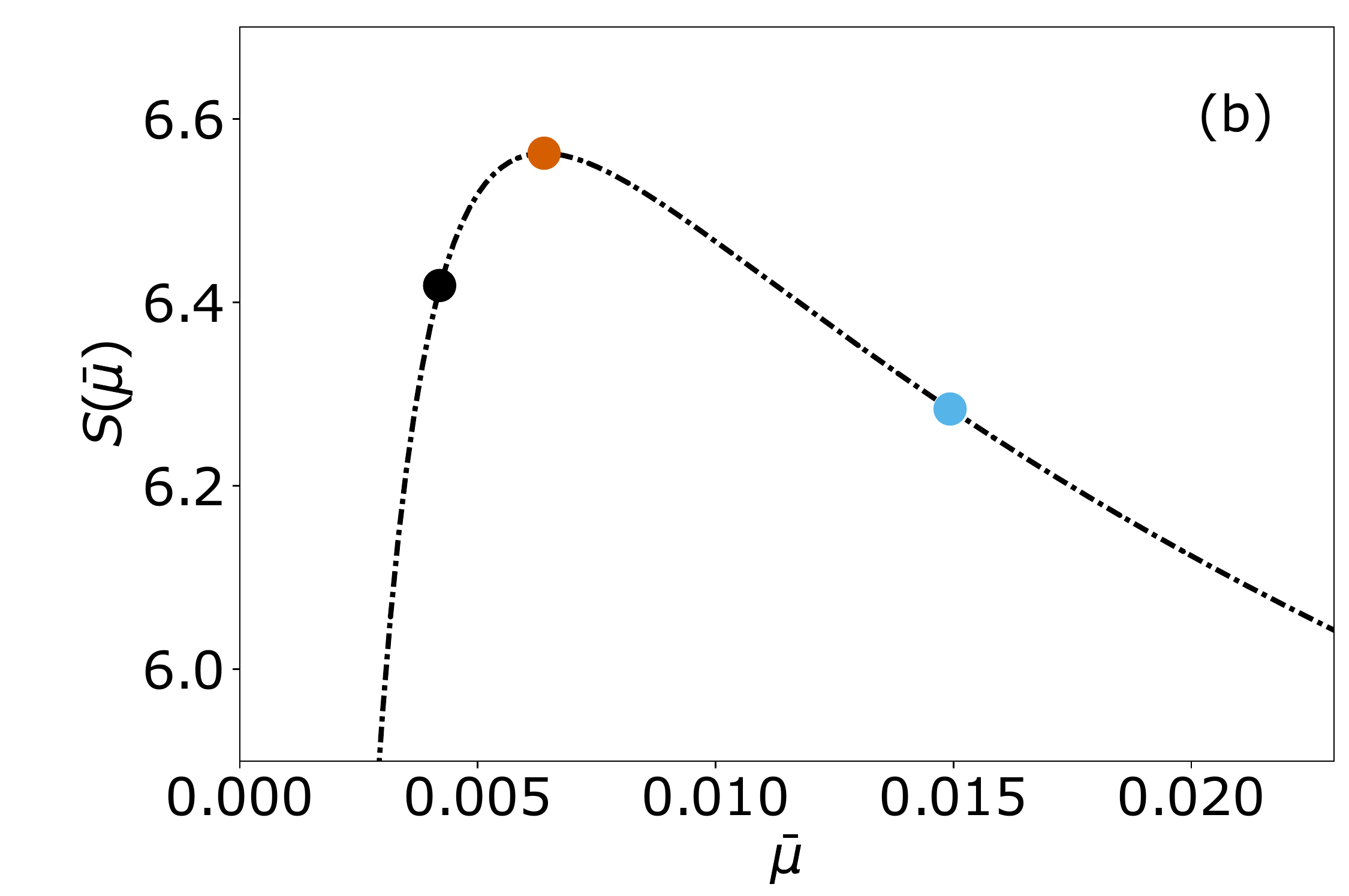}
		\caption{(color online) (a) Stationary probability distributions of macrostates for three selected values of the average mutation $\bar\mu$; (b) Shannon entropy as a function of the average mutation rate for the Moran model on a fully connected network with $N=100$ nodes and asymmetric mutation rates. The magnitude of the asymmetry parameter $\Delta \equiv \mu_+ - \mu_-$ was set to $0.0012$. The dotted line in panel (a) marks the average value of $m$ of the symmetric case $\Delta=0$. Colored points on the Shannon entropy function ((b), dashed line) correspond to the curves on panel (a).}
		\label{fig::rho_shannon_asym}
	\end{figure}

	We first focus on analyzing a fully connected network with a particular asymmetry parameter value, $\Delta$. In Fig. \ref{fig::rho_shannon_asym}(a), we show the stationary distributions for three values of $\bar{\mu}$. Contrary to the symmetric case (Fig. \ref{fig::FC_plots}(a)), the transition (as $\mu_c$ increases) between the low-diversity stationary distribution (black line, Fig.  \ref{fig::rho_shannon_asym}(a)) and the high-diversity stationary distribution (bell-shaped blue curve, Fig. \ref{fig::rho_shannon_asym}(a)) is not characterized by a uniform distribution, as the asymmetry skews the distribution for all values of $\bar{\mu}$. The Shannon entropy curve, however, provides an unambiguous identification of the transition point (Fig.  \ref{fig::rho_shannon_asym}(b)). In this case, the parameter $\bar{\mu}_c$ whose corresponding stationary distribution maximizes the Shannon entropy is found to be approximately $\bar{\mu}_c=0.0064$, higher than the value $\bar{\mu}_c=0.005$ corresponding to the symmetric case.  Moreover, the stationary distribution corresponding to the transition point $\bar{\mu}_c=0.0064$ (orange line, Fig. \ref{fig::rho_shannon_asym}(a)) seems to nicely interpolate between the low- and high-diversity phases (black and blue curves, respectively).
	
\begin{figure*}[!htpb]
		\centering
		\includegraphics[width=0.32\textwidth]{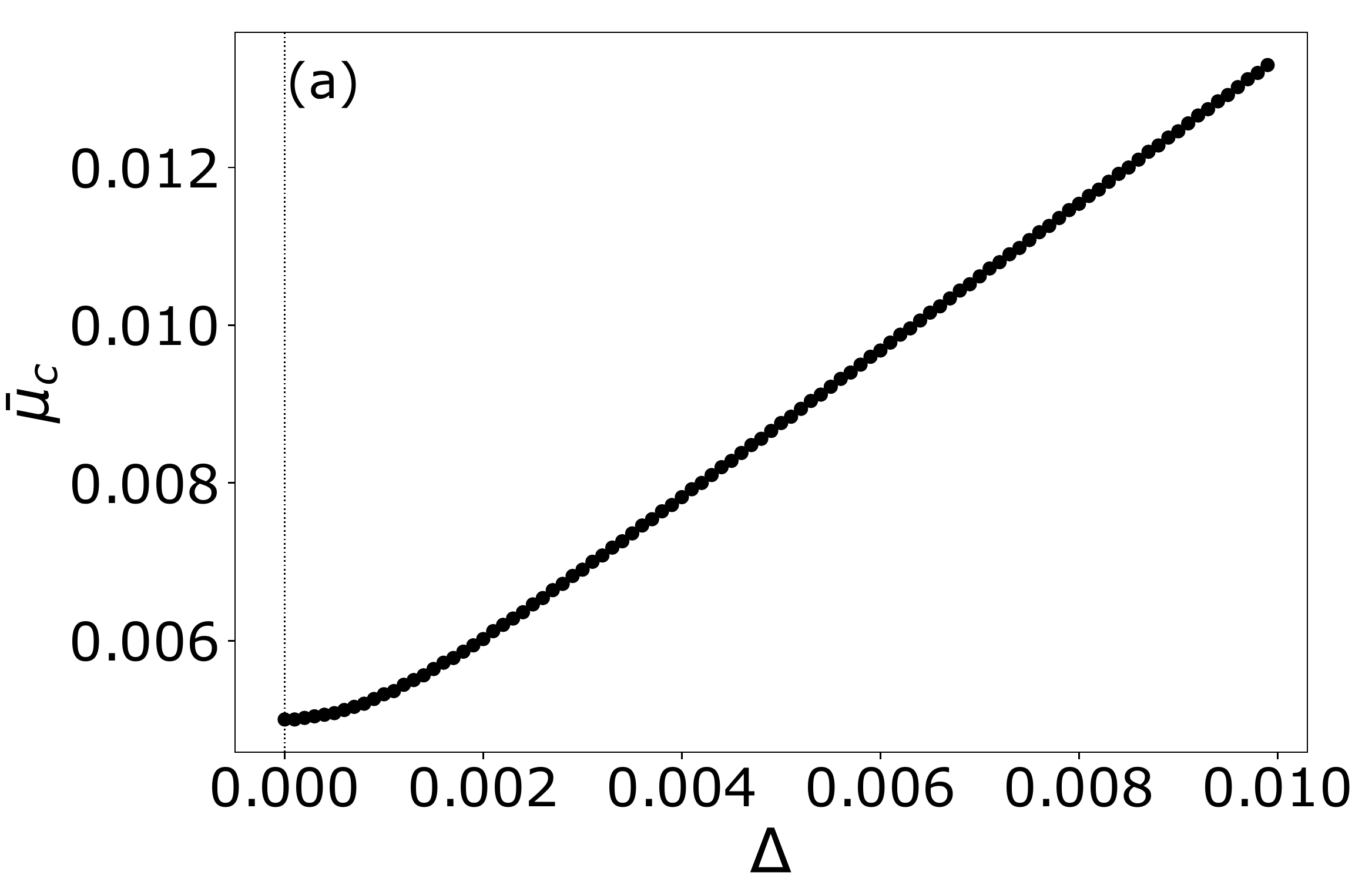}
		\includegraphics[width=0.32\textwidth]{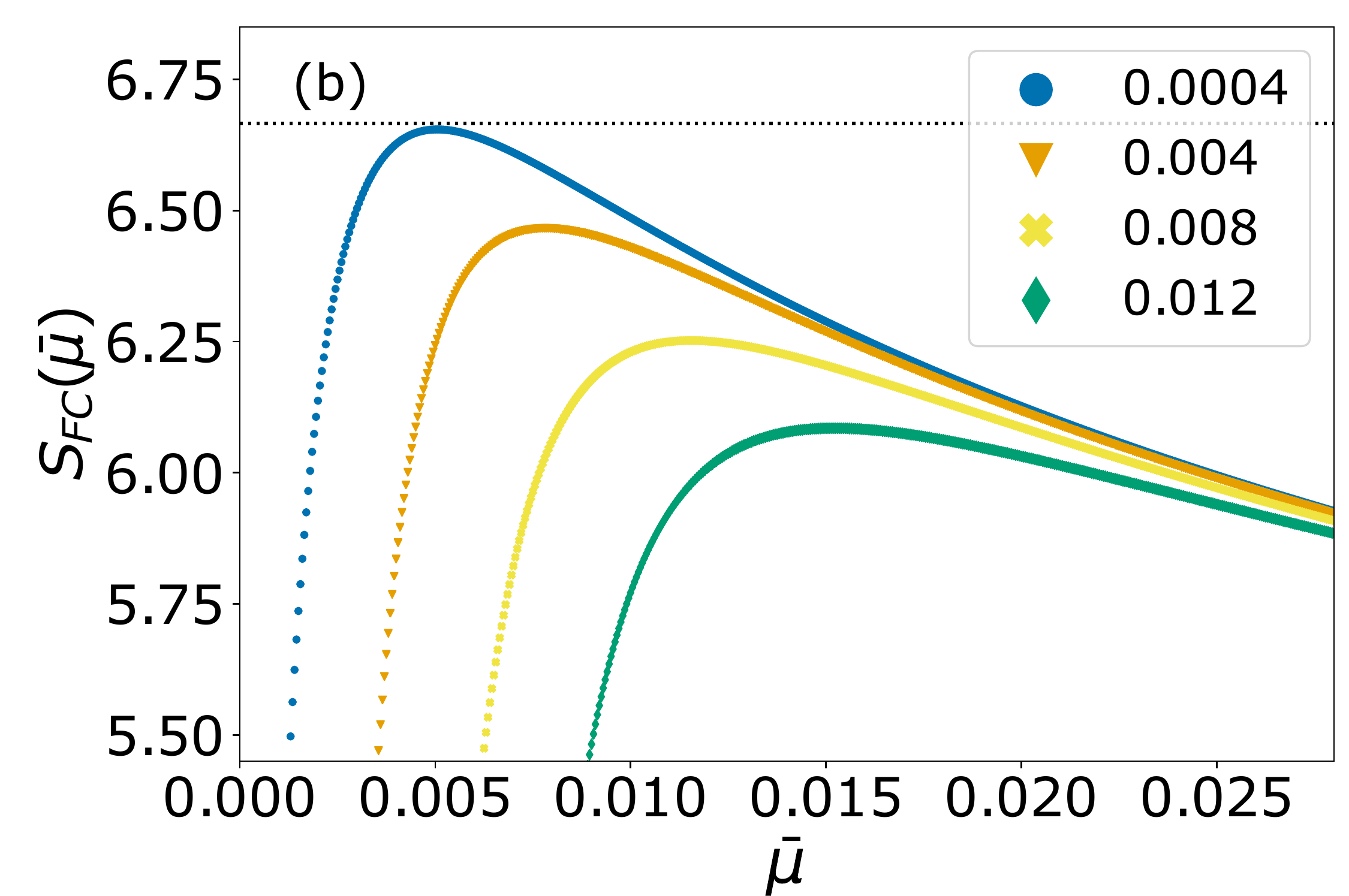}
		\includegraphics[width=0.32\textwidth]{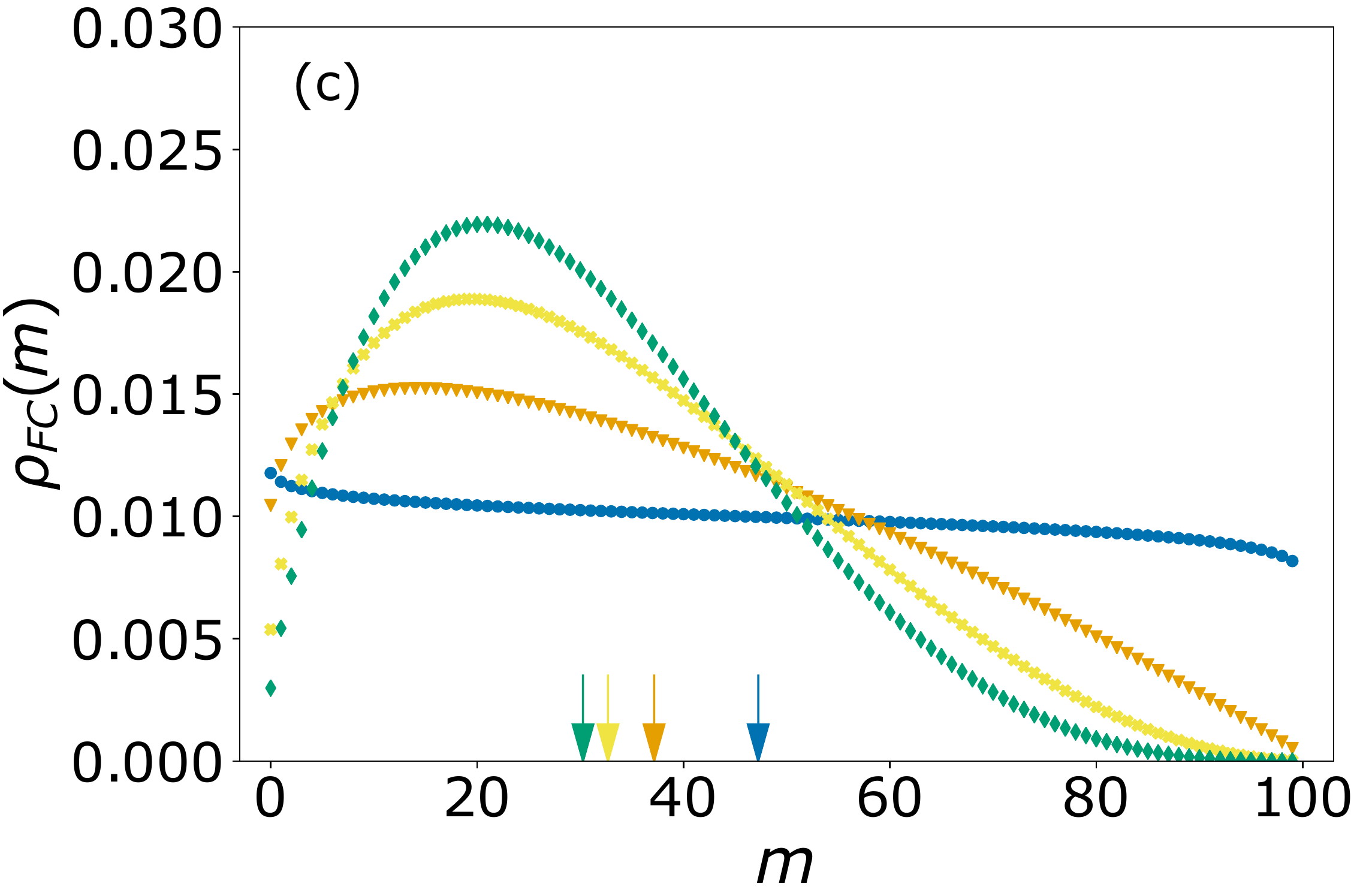}
		\caption{(color online) (a) Relation between asymmetry ($\Delta$) and transition point ($\mu_c$) for a fully connected network with $N=100$. (b) Shannon entropy curves as a function of the average mutation rate corresponding to four values of $\Delta$. (c)   Stationary probability distributions of macrostates at the transition points. The distribution curves correspond to the same asymmetry parameters $\Delta$ specified in panel (b). Arrows denote the mean value of the corresponding distributions.}
		\label{fig::mu_delta}
	\end{figure*}

	To further understand the effect of asymmetry on the low-high diversity transition point, plots of the transition point $\bar{\mu}_c$ as a function of the asymmetry parameter $\Delta$ are shown in Fig. \ref{fig::mu_delta}(a). We see that the transition point $\bar{\mu}_c$ increases as the magnitude of the asymmetry $\Delta$ increases. For $\Delta \approx 0$, the derivative of $\bar{\mu}_c$ with respect to $\Delta$ tends to zero and, therefore, the transition point does not change significantly for small values of the asymmetry parameter. Fig. \ref{fig::mu_delta}(b) shows the Shannon entropy as a function of $\bar{\mu}_c$ for four different values of $\Delta$. We see that increasing the degree of asymmetry displaces the peak position of the entropy function to higher values of $\bar{\mu}_c$ -- consistent with Fig. \ref{fig::mu_delta}(a) -- as well as to smaller values of $S(\bar{\mu}_c)$. Fig. \ref{fig::mu_delta}(c) shows the equilibrium distributions corresponding to the transition points identified in Figure \ref{fig::mu_delta}(b). Overall, these results suggest that the equilibrium distribution, entropy function, transition point $\bar{\mu}_c$, and its associated maximum entropy values $S(c)$ may vary significantly with the asymmetric parameter $\Delta$.
	
	\subsection{The effect of network randomness on the low-high diversity transition}
	\label{subsec::SW}
	
\begin{figure}[!htpb]
		\centering
		\includegraphics[width=0.9\columnwidth]{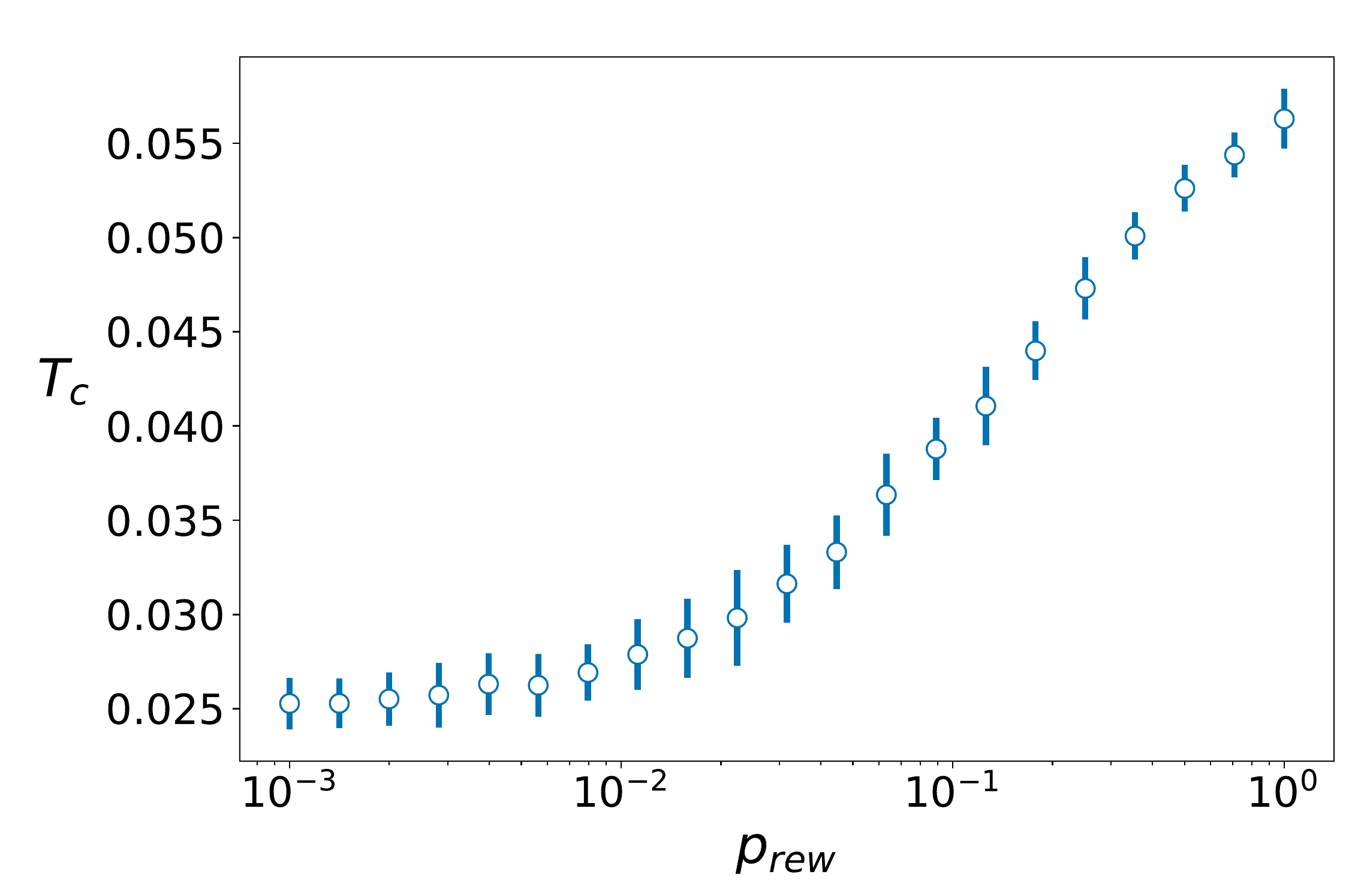}		\caption{(color online) Transition points $T_c$ as a function of the rewiring probability $p_{rew}$ for the Voter model on small-world networks. For $p_{rew}=0$ the network is a ring with $N=100$ nodes and $k=6$. The transition points $T_c$ are averages over $50$ simulations for each $p_{rew}$, generating a different randomly rewired network in each case. Error bars correspond to one standard deviation computed over 50 realizations with different randomly generated networks.}
		\label{fig::SW}
	\end{figure}

	While regular networks serve as useful models for complex systems, many real networks are neither completely regular nor completely random. It is therefore interesting to analyze the impact of network randomness on the transition point. To this end, we consider small-world networks where randomness is controlled by a probability $p_{rew}$ of randomly rewiring each edge of an initially ring network \cite{watts1998collective}. A distinctive property of small world networks is that the average path length between nodes rapidly decreases with $p_{rew}$, whereas the clustering coefficient remains virtually unchanged. For simplicity, we consider the Voter model with symmetric parameters, and an initial ring network with $N = 100$ nodes and degree $k = 6$. Fig. \ref{fig::SW} shows the transition point $T_c$ as a function of $p_{rew}$. For $p_{rew}=0$ we obtain $T_c=0.025 \pm 0.001$, which is consistent with the results for the ring network (Fig. \ref{fig::rho_plts-ring_lattice}(a)). On the other hand, when $p_{rew}=1$ the underlying network becomes completely random, and we obtain $T_c=0.058\pm 0.001$ close to the mean-field approximation $T_c \approx \frac{<k>}{(N-1)} = 0.061$.
	
	We see in Fig. \ref{fig::SW} that the transition point $T_c$ increases as the degree of randomness in the network increases. We note that  a similar effect was observed in previous studies, demonstrating an increase in the critical transition $T_c$ when considering randomness in the external influence parameters \cite{ramos2018opinion}. We can interpret these results as follows: increasing the randomness in the network reduces the impact of the external influences by increasing the critical threshold $T_c$ required for the transition. The  low-high diversity transition in this case can be directly associated to changes in the characteristic path length of the network, which decreases as more links are rewired and the degree of randomness increases. This increased randomness enables opinions (or alleles) to spread more easily throughout the network, thus leading faster to low diversity states. This is consistent with the findings that shorter path lengths facilitate imitation by other nodes, leading to the rapid spread of opinions and promoting consensus \cite{braha2017,dornelas2018}. Although our discussion here focuses on the Voter model, completely analogous considerations can be made for the Moran model.

	\section{Discussion}\label{sec::discussion}
	
	Understanding the factors that influence the transition between consensus (low diversity) and plurality (high diversity) is fundamental in a variety of social and biological contexts. Here we investigate this transition for the Voter and Moran models on networks, which are subjected to external influencers or mutation rates, respectively. Previous work derived a mean-field approximation for the equilibrium density of macrostates for regular networks \cite{mobilia2007role,schneider2016mutation}. However, this work failed to adequately provide a reasonable definition for the transition point, which can be applied more broadly to arbitrary networks with asymmetric external influence parameters or mutation rates. Here we propose a new information-theoretic method for determining this transitional state by utilizing the Shannon entropy corresponding to the stationary probability distribution of macrostates. The transition point is simply defined as the set of parameters whose corresponding stationary distribution maximizes the Shannon entropy.
	
	Phase transitions for macrostates are found in similar versions of the Voter model \cite{mobilia2007role, Celia2020,mobilia2015nonlinear,sood2005voter}. One way to study these transitions is by analogy with ferromagnetic systems, using traditional tools from statistical mechanics. For the model studied here, the analogue of the magnetization is always a smooth function of the external parameters and no sharp transition occurs between the low and high diversity phases for finite-size systems. In the limit $N \rightarrow \infty$, the critical point goes to zero, indicating that any external perturbation leads to some degree of diversity. An alternative way to characterize the transition between the phases is by looking at the point where the distribution of macrostates changes from unimodal (high diversity) to bimodal (low diversity) \cite{khalil2018zealots}. This definition does not coincide with the one proposed here, as can be seen from Fig. \ref{fig::rho_shannon_asym}(a) where the maximum of the Shannon entropy is for $\bar\mu = 0.0064$, which still represents a unimodal distribution. A more dramatic example occurs for the star network \cite{moreira2015binary} where the high to low diversity transition is rather different. In this case the single peak of distribution of macrostates in the unimodal regime splits in two as the external influence is decreased (see figure 3 of \cite{moreira2015binary}). For $N=100$ nodes and $T=N_0=N_1$, the point of highest entropy is achieved for $T=0.32$ whereas the point where the peak splits in two is $T=2.8$, showing that the two metrics of transition are indeed measuring different properties of the system.
	
	Our information-theoretic approach offers a natural way to define the transition point for finite-size systems with arbitrary network topology and asymmetric external influences. We demonstrated the usefulness of this approach  by extensive numerical simulations on both homogeneous and heterogeneous networks. These results show that the low-high diversity transition is controlled by a variety of factors including the network topology, average node degree, degree of asymmetry in external influence parameters, and level of randomness in network topology. In particular, we report an intriguing relationship between the characteristic path length of small-world networks and their corresponding transition points.
	
	In biological systems low diversity (and consensus) is associated to monomorphism, and high diversity (and coexistence of opinions) is associated to polymorphism. In non-structured populations (fully-connected network), the transition point corresponds to the mutation rate of $1/2N$ \cite{chinellato2015dynamical}, as predicted by the mutation-drift balance \cite{gillespie2004population}. For populations structured in regular network topologies, the mean-field approximation for the transition point in the Voter model is $T_c(k)=k/(N-1)$. This maps to $\mu_c=1/(2N)$ in the Moran model, which is identical to the value for fully connected populations. However, in biological systems polymorphisms can be promoted by structured populations, especially  those that are structured by geography \cite{hutchison1999, planes2002,leal2019dispersal}, decreasing the critical mutation rate required for the low-high diversity transition. This expectation is verified by our simulations, which show that the critical mutation rate $\mu_c$ (detected by the Shannon entropy criterion) is dependent on the average node degree of the network, counter to the prediction of the mean-field theory. More specifically, we find that the critical mutation rate for the Moran model is well described by $\mu_c=\frac{ak+b}{2(a+1)k+b}$ where $a$ and $b$ are fitted parameters (see Fig. \ref{fig::results_degree}). 
	
	We also demonstrated that the degree of asymmetry in external influence parameters leads to skewed stationary probability distributions, emphasizing the inadequacy of using the uniform probability distribution as a criterion for the transition. Skewness of the probability stationary distributions is particularly  relevant in small natural populations, which can remain monomorphic for long periods  before a mutant appears. 
	
	Our choice of the Shannon entropy as a diversity measure, as we noted before, is not unique. Thus, comparing the various measures with respect to their ability to identify the high-low diversity transition point is of interest in itself and will be addressed in future work. Another line of investigation concerns different network topologies (particularly modular networks), and their effect on the transition. For non-regular networks there is no exact mapping between the Voter and Moran models. It would thus be interesting to understand how their dynamics might diverge for more complex network topologies. This could help to elucidate the interplay between geographic population structures and the critical mutation rate required for the existence of polymorphisms.

	\begin{acknowledgments} 
		This work was partly funded by the Brazilian agencies CNPq, grants \#301082/2019-7 (MAMA) and \#132438/2017-8 (GDF); FAPESP, grants \#2016/01343-7, \#2019/20271-5 (MAMA) and \#2020-13957-5 (GDF); and Coordenação de Aperfeiçoamento de Pessoal de Nível Superior -- Brasil (CAPES) -- Finance Code 001 (FMDM).
	\end{acknowledgments}

	\appendix
	\begin{widetext}
		
		\section{Master equations}
		\label{app:maseqn}
		
		In this appendix we briefly review the derivation of the master equations for the Moran and Voter models. We call $x_i \in \{0,1\}$ the state of individual $i$ and ${\bf x}= \{x_1,...,x_j,...,x_N\}$ the state of the population. We recall that, in both models, the state of a single individual is allowed to change at each time step. This implies that there are only two ways to reach a given state ${\bf x}$ at time $t+1$: (i) either the system is already at ${\bf x}$ at time $t$ or; (ii) the state at time $t$ differs from ${\bf x}$ by a single individual. We define the state that differs from ${\bf x}$ by the state of individual $j$ as ${\bf x}^j= \{x_1,...,1-x_j,...,x_N\}$. The master equation can now be written as
		\begin{equation}
			P_{t+1}({\bf x})=P_t({\bf x})\Pi({\bf x}\rightarrow {\bf x})+\sum_{i=1}^N P_t({\bf x}^i)\Pi({\bf x}^i\rightarrow {\bf x}),
			\label{master_general}
		\end{equation}
		where $P_t({\bf x})$ indicates the probability of finding the population in state ${\bf x}$ at time $t$ and $\Pi({\bf a}\rightarrow{\bf b})$ is the transition probability from state ${\bf a}$ to state ${\bf b}$.
		
		In the Moran model, a step of the dynamics consists in selecting a random (focal) individual and replacing it by its offspring with another randomly selected (partner) individual. The offspring may inherit the state (allele) of the focal individual or the partner with equal probability.  The inherited allele also has a probability to mutate. Here we shall exchange the word `individual' by 'network node' and also use the term `copy the state' for 'inherit the allele'.
		
		The transition probability $\Pi({\bf x}\rightarrow {\bf x})$ of remaining in the same state is given by either copying the state of the focal node (which then cannot mutate) or the state of a mating partner with the same state of the focal node (without mutation) or the opposite state with mutation. The contribution to the probability for copying the focal node is
		\begin{align}
			\nonumber\Pi_{f}({\bf x}\rightarrow {\bf x}) = &\sum_{i=1}^N \frac{1}{N}\frac{1}{2}[x_i(1-\mu_-)+(1-x_i)(1-\mu_+)],\\
			&= 1-\frac{1}{2N}\sum_{i=1}^{N}\left[1+x_i\mu_- +(1-x_i)\mu_+\right].
		\end{align}
		where $\mu_-$ is the mutation probability from state 1 to 0 and $\mu_+$ from 0 to 1. This represents the probability of picking individual $i$ ($1/N$) times the probability of copying its state ($1/2$), summed over all individuals.
		
		The contribution coming from copying the mating partner follows the same idea but now involves the adjacency matrix to select connected neighbors:
		\begin{align}
			\nonumber \Pi_m({\bf x}\rightarrow {\bf x})=&\frac{1}{2N}\sum_{i=1}^N \frac{1}{k_i}\left\{ \sum_{j=1}^{N} A_{ij}\lvert 1-x_i-x_j\rvert  [ x_i(1 - \mu_-) \right.  + (1-x_i)(1-\mu_+)] \Bigg\} \\
			\nonumber+&\sum_{j=1}^{N} A_{ij}\lvert x_i-x_j\rvert \left[x_i\mu_+ \right.  \left. + (1-x_i)\mu_-\right]  \nonumber
			\\  =& \frac{1}{2N}\sum_{i=1}^{N}\frac{1}{k_i}\sum_{j=1}^{k_i}\left\{A_{ij}\lvert 1-x_i-x_j\rvert (1-2\bar{\mu}) \right. \left. + k_i[x_i\mu_+(1-x_i)\mu_-]\right\},
			\label{p2}
		\end{align}
		where we defined $\bar{\mu}=(\mu_+ + \mu_-)/2$.
		
		The transition probability $\Pi({\bf x}^i \rightarrow {\bf x})$ also has two contributions. In both cases the node whose state differs has to be selected ($1/N$). If the state of focal node is copied, then is has to mutate:
		\begin{equation}
			\Pi_f({\bf x}^i\rightarrow {\bf x})=\frac{1}{2N}\left\{ x_i^i\mu_-+(1-x_i^i)\mu_+ \right\}.
		\end{equation}
		The second possibility again comes from copying the mating partner and allowing or restricting mutations accordingly:
		\begin{align}
			\nonumber \Pi_m({\bf x}^i\rightarrow {\bf x})=&\frac{1}{2N}\frac{1}{k_i}\Bigg\{\sum_{j=1}^{N}A_{ij}\lvert x_i^i-x_j\rvert \left[x_i^i(1-\mu_-)
			+ (1-x_i^i)(1-\mu_+)\right] \\
			\nonumber  +& A_{ij}\lvert 1-x_i^i-x_j\rvert
			\left[ x_i^i\mu_+ + (1- x_i^i)\mu_- \right] \Bigg\}  
			\\ \nonumber =& \frac{1}{2N}\frac{1}{k_i}\Bigg\{\sum_{j=1}^{N}A_{ij}\lvert 1-x_i-x_j\rvert \left[ (1-x_i)(1-2\bar{\mu}) +x_i(1-2\bar{\mu}) \right] \\
			\nonumber  +& k_i\left[(1-x_i)\mu_+ + x_i\mu_-\right]\Bigg\} .
		\end{align}
		
		Putting these terms together and simplifying we find the master equation
		\begin{align}
			P_{t+1}(x) &= P_t(x)  + \notag \\ 
			&+ P_t(x) \frac{(1-2\bar{\mu})}{2N}  \sum_i
			\frac{1}{k_i} \left\{ \sum_j  A_{ij} |1-x_i-x_j| -k_i  - \frac{ 2  \mu_+ k_i}{1-2\bar{\mu}} x_i - \frac{2 \mu_- k_i}{1-2\bar{\mu}}(1-x_i)  \right\} + \notag \\ 
			& + \frac{(1-2\bar{\mu})}{2N} \sum_i \frac{P_t(x^i)}{k_i} \left\{
			\sum_j A_{ij}|1-x_i-x_j| + \frac{2\mu_+ k_i}{1-2\bar{\mu}}(1-x_i) +
			\frac{2\mu_- k_i}{1-2\bar{\mu}}x_i \right\} .
			\label{master_moran}
		\end{align}
		
		A similar procedure can be applied for the Voter model dynamics. An important difference appears from the fact that the external influence, given by nodes $N_0$ and $N_1$, only affects the system when the node copies the state of the partner. In the language of Moran this would be as if `mutations' occurred only if the state of the mating partner is selected. Another difference comes from the parameter $p$ which defines the chance of the system remaining unchanged at that time step. This parameter only changes the time scale for reaching the stationary probability distribution, not the distribution itself. The master equation for the Voter model is given by:
		\begin{align}
			\nonumber P_{t+1}({\bf x}) &= P_{t}({\bf x})\ +\\ \notag
			&+\frac{(1-p)}{N}P_t({\bf x})\sum_{i=1}^{N} \frac{1}{k_i+N_0+N_1}\left\{\sum_{j=1}^{N-1}A_{ij}|1-x_i-x_j| - k_i  - N_0 x_i - N_1(1-x_i) \right\} + \\ & +\frac{(1-p)}{N}\sum_{i=1}^{N}\frac{P_t({\bf x}^i)}{k_i+N_0+N_1}\left\{\sum_{j=1}^{N-1} A_{ij}|1-x_i-x_j| \linebreak + N_0(1-x_i) + N_1 x_i  \right\}.
			\label{master_voter}
		\end{align}
		
		The solution for the fully connected case of the Voter model with external influencers can be found in \cite{chinellato2015dynamical}. The analytic stationary probability distribution is given by:
		
		\begin{equation}
			\rho_{FC}^{N_0,N_1}(m) =  A_{N}^{N_0,N_1} ~ \frac{\Gamma(N_1+m)~\Gamma(N + N_0 - m)}
			{\Gamma(N-m+1)~\Gamma(m+1)},
			\label{probn}
		\end{equation}
		where
		\begin{equation*}
			A_N^{N_0,N_1} = \frac{\Gamma(N+1)~\Gamma(N_0+N_1)}{\Gamma(N + N_0 + N_1)
				~\Gamma(N_1)~\Gamma(N_0)}.
		\end{equation*}
		The gamma functions allow the extension of the model to real values of $N_0$ and $N_1$, representing continuous ranges of external influences.
		
	\end{widetext}

\end{document}